\documentclass[reqno]{amsart}
\usepackage{amsfonts,amsmath,amssymb,bbm}
\usepackage{verbatim}
\usepackage[ngerman,american]{babel}
\usepackage{url}
\usepackage{bbm}
\usepackage{color}

\definecolor{Red}{rgb}{1,0,0}
\definecolor{Blue}{rgb}{0,0,1}
\definecolor{Green}{rgb}{0,1,0}
\definecolor{Yellow}{rgb}{1,1,0}
\definecolor{Cyan}{rgb}{0,1,1}
\definecolor{Magenta}{rgb}{1,0,1}
\definecolor{Orange}{rgb}{1,.5,0}
\definecolor{Violet}{rgb}{.5,0,.5}
\definecolor{Purple}{rgb}{.75,0,.25}
\definecolor{Brown}{rgb}{.75,.5,.25}
\definecolor{Grey}{rgb}{.5,.5,.5}

\newcommand{\etalchar}[1]{$^{#1}$}

\providecommand{\MR}{\relax\ifhmode\unskip\space\fi MR }
 \providecommand{\href}[2]{#2}

\def\bes{\begin{subarray}{c} }
\def\es{\end{subarray} }

\def\sump{\!\!\mathop{\,\,{\sum}^{\,\prime}}}
\def\sumpp{\mathop{{\sum}^{\,\prime\prime}}}

\def\RE{\Re\text{\rm e}}
\def\IM{\Im\text{\rm m}}
\def\R{{\mathbb R}}

\def\PR{{\mathbb P}}
\def\EX{{\mathbb E}}
\newcommand{\eq}{\begin{equation}}
\newcommand{\en}{\end{equation}}
\newcommand{\eqa}{\begin{eqnarray}}
\newcommand{\ena}{\end{eqnarray}}
\newcommand{\eqas}{\begin{eqnarray*}}
\newcommand{\enas}{\end{eqnarray*}}
\newcommand{\ra}{ {\rightarrow} }

\newcommand{\npp}{\textsc{Npp}}

\newcommand{\abs}[1]{\ensuremath{\left\vert #1 \right\vert}}
\providecommand{\binom}[2]{\left(#1\atop#2\right)} \makeatletter
\def\erf{\mathop{\operator@font erf}\nolimits}
\def\erfc{\mathop{\operator@font erfc}\nolimits}
\def\argmax{\mathop{\operator@font argmax}\nolimits}
\def\sinc{\mathop{\operator@font sinc}\nolimits}
\def\rect{\mathop{\operator@font rect}\nolimits}
\def\poly{\mathop{\operator@font poly}\nolimits}
\@ifundefined{widetext}{}{}
\makeatother

\def\bss{\boldsymbol{\sigma}}
\def\bsd{{\boldsymbol{\delta}}}

\newtheorem{theorem}{Theorem}[section]
\newtheorem{proposition}[theorem]{Proposition}
\newtheorem{lemma}[theorem]{Lemma}

\theoremstyle{definition}

\theoremstyle{remark}
\newtheorem{remark}[theorem]{Remark}

\numberwithin{equation}{section}

\newcommand{\rhat}{\hat{\rho}}

\begin{document}

\title[Proof of the local REM conjecture for number partitioning II]
{Proof of the local REM conjecture for number partitioning II:
growing energy scales}

\author{Christian Borgs$^1$, Jennifer Chayes$^1$, Stephan Mertens$^2$,
        Chandra Nair$^1$}

\address{$^1$Microsoft Research, One Microsoft Way, Redmond, WA 98052}

\address{\selectlanguage{ngerman}{$^2$Inst.\ f.\ Theor.\ Physik,
    Otto-von-Guericke Universit"at, PF~4120, 39016 Magdeburg, Germany}}



\date{August 23, 2005}


\begin{abstract}
We continue our analysis of the number partitioning problem with $n$
weights chosen i.i.d.~from some fixed probability distribution with
density $\rho$. In Part I of this work, we established the so-called
local REM conjecture of Bauke, Franz and Mertens.  Namely, we showed
that, as $n \to \infty$, the suitably rescaled energy spectrum above
some {\it fixed} scale $\alpha$ tends to a Poisson process with
density one, and the partitions corresponding to these energies
become asymptotically uncorrelated. In this part, we analyze the
number partitioning problem for energy scales $\alpha_n$ that {\it
grow with $n$}, and show that the local REM conjecture
holds as long as $n^{-1/4}\alpha_n \to 0$,  and fails if
$\alpha_n$ grows like $\kappa n^{1/4}$ with $\kappa>0$.

We also consider the SK-spin glass model, and show that
it has an analogous threshold: the local REM conjecture
holds for energies of order
$o(n)$, and fails if the energies grow like
$\kappa n$ with $\kappa >0$.
\end{abstract}

\maketitle

\section{Introduction}
\label{sec:intro}

\subsection{Number Partitioning}
In this paper we continue the study of the energy spectrum
of the number partition problem (\npp) with randomly
chosen weights.
We refer the reader to \cite{part1} for a detailed
motivation of this study,
but for completeness, we repeat the main definitions.

We consider random instances of the {\npp} with
weights $X_1,\dots,X_n\in\R$ taken to be independently and
identically distributed  according to some
density $\rho(X)$
with finite second moment (for simplicity of notation, we will
choose the second moment to be one).
Given these weights,
one seeks a partition of
these numbers into two subsets such that the sum of numbers in one
subset is as close as possible to the sum of numbers in the other
subset.
Each of the $2^n$ partitions can be encoded as
 $\bss \in  \{-1,+1\}^n$, where $\sigma_i=1$ if
$X_i$ is put in one subset and $\sigma_i=-1$ if $X_i$ is put in the
other subset; in the
physics literature, such partitions $\bss$ are identified with {\em Ising
spin configurations}.  The cost function to be minimized over all spin
configurations $\bss$ is  the {\em energy}
\begin{equation}
  \label{ene}
  E(\bss) = \frac{1}{\sqrt{n}}\left|\sum_{i=1}^n \sigma_i X_i \right|,
\end{equation}
where, as in \cite{part1}, we have inserted a factor
$1/\sqrt{n}$
to simplify the
equations in the rest of the paper.

Note that this scaling implies that the {\em typical}
energies are of order one, and the maximal energies are of
order $\sqrt n$. Indeed, if $\bss$ is chosen uniformly at random and
$X_1,\dots,X_n$ are i.i.d. with second moment one, the random
variable $n^{-1/2}\sum_i\sigma_iX_i$ converges to a standard normal
as $n\to\infty$, implying in particular that for a typical
configuration, $E(\bss)$ is  of order one. The maximal energy,
on the other hand, is equal to $n^{-1/2} \sum_i|X_i|$. By the law of
large numbers, this implies that the maximal energy is
asymptotically equal to $\sqrt n$ times the expectation of $|X|$.

As usual, the correlation between two
different partitions $\bss$ and $\tilde\bss$ is measured
by the \emph{overlap} between $\bss$ and $\tilde\bss$,
defined as
\begin{equation}
\label{overlap} q(\bss,\tilde\bss)= \frac 1n
\sum_{k=1}^n\sigma_k\tilde\sigma_k .
\end{equation}
Note that the spin configurations $\bss$ and $-\bss$ correspond to
the same partition and therefore of course have the same energy.
Thus there are $N = 2^{n-1}$ distinct partitions and
(with probability one) also $N$
distinct energies.  The {\em energy spectrum} is the
sorted increasing sequence $E_1,...,E_N$ of the energy values
corresponding to these $N$ distinct partitions.
Taking into account that, for each
$i$, there are  two configurations
 with energy $E_i$, we
define $\bss^{(i)}$ to be the random
variable which is equal to one of these two
configurations with probability $1/2$, and equal
to the other with probability $1/2$.  Then the
overlap between the configurations corresponding to
$i^{\text{th}}$
and $j^{\text{th}}$ energies is the random variable
$q(\bss^{(i)},\bss^{(j)})$.

As noted in \cite{part1},  neither the distribution of the energies,
nor the distribution of the overlaps changes if one replaces
the density $\rho(X)$ by the symmetrized density
$\frac 12(\rho(X)+\rho(-X))$.  We may therefore assume without loss
of generality that $\rho(X)=\rho(-X)$.  Under this assumption,
it is easy to see that the energies $E(\bss)$ for the different
configurations $\bss$ are identically distributed. Let us stress,
however, that these energies are \emph{not} independently
distributed; the
energies between
different configurations
are \emph{correlated} random variables.  Indeed, there are
$N = 2^{n-1}$ energies, $E_1, \dots E_n$, constructed from
only $n$ independent variables $X_1, \dots X_n$.

Consider now a very simple model, the so-called random energy model
(REM) first introduced by Derrida \cite{derrida:81}
in a different
context.  The defining property of the REM is that  $N$
energies $E(\bss)$ are taken to be
\emph{independent}, identically distributed
random variables.  In the REM,
the asymptotic energy spectrum for large  $N$ can be easily
determined with the help of large order statistics: if
the energies are ordered in increasing order and
$\alpha\geq 0$ is any fixed energy scale, the suitably rescaled
energy spectrum
above $\alpha$ converges to a Poisson process.
More precisely, if the distribution of $E(\bss)$ has a non-vanishing,
continuous
density $g(\alpha)$ at $\alpha$ and $E_{r+1}$ is the first energy above
$\alpha$, then the rescaled energies
$(E_{r+1}-\alpha)Ng(\alpha)$, $(E_{r+2}-\alpha)Ng(\alpha)$,
$\dots$ converge to a Poisson process with density one.

In spite of the correlations between
the energies $E(\bss)$ in the $\npp$,  it
had been conjectured \cite{mertens:00a,rem1} that as $n\to\infty$,
the energy spectrum above any fixed energy $\alpha$ behaves
asymptotically like the energy spectrum of the
REM, in the sense that the suitably rescaled spectrum above
$\alpha$ again becomes a Poisson process.
In \cite{rem1} it was also
conjectured that the overlaps corresponding to
adjacent energies are
asymptotically uncorrelated, so that
the suitably normalized overlaps converges to a standard normal.
These two, at first sight highly speculative, claims were collectively
called the {\em local REM conjecture}.  This conjecture
was supported by detailed simulations.

In Part I of this paper \cite{part1}, we proved the local REM
conjecture for the \npp\ with a distribution $\rho$ that has finite
second moment and lies in ${L}^{1+\epsilon}$ for
some $\epsilon>0$. More precisely, under these conditions, we proved
that for all $i\neq j$, the suitably normalized overlap between the
configurations corresponding to the $i^{\text{th}}$ and
$j^{\text{th}}$ energy above $\alpha$ becomes asymptotically normal,
and that the rescaled energies $(E_{r+1}-\alpha)\xi_n^{-1}$,
$(E_{r+2}-\alpha)\xi_n^{-1}$, $\dots$ with rescaling factor
\begin{equation}
\label{xi-n}
\xi_n=\sqrt{\frac{\pi }{2}} 2^{-(n-1)}
e^{\frac{\alpha^2}{2}}
\end{equation}
converge to a Poisson process with density one.
Recalling that the normalization
in \eqref{ene} corresponds to typical energies of order one,
this this establishes the local REM conjecture
for \emph{typical} energies.

In \cite{rem1}, the authors expressed the belief that the weak
convergence of the rescaled energies to a Poisson process should
extend to values of $\alpha$ that grow slowly enough with $n$,
although computational limitations prevented them from supporting
this stronger claim by simulations.
At first, one might think that the local REM conjecture
could hold for $\alpha = o(\sqrt n)$.  Indeed,
since the maximal energy
is of order $\sqrt n$, it is clear that the conjecture is
false for $\alpha = c\sqrt n$ with large enough
$c$.  But if this were
the only obstruction,
then one might hope that the conjecture could hold up
to  $\alpha = o(\sqrt n)$. As we will see in this paper,
this is \emph{not the case}; the conjecture will only hold for
$\alpha=o(n^{1/4})$.

Before stating this precisely, let us be a little
more careful with the scaling of the energy spectrum. Note that the
scaling factor \eqref{xi-n} is equal to $[N g(\alpha)]^{-1}$, where
$N=2^{n-1}$ is the number of energies, and $g(\alpha)=\sqrt{
2/\pi}e^{-\alpha^2/2}$ is the density of the absolute value of a
standard normal, in accordance with the expected asymptotic density
of $E(\bss)$ according to the local limit theorem.  But it is well
know that, in general, the local limit theorem does not hold in the
tails of the distribution. For growing $\alpha_n$, the
REM conjecture should therefore be stated with a scaling factor that
is equal to the inverse of $2^{n-1}$ multiplied by the
density of the energy $E(\bss)$ at $\alpha$. We call the
REM conjecture with this scaling the \emph{modified} REM conjecture.
It is this modified REM conjecture that one might naively expect
to hold for $\alpha = o(\sqrt n)$.

It turns out, however, that at least for the {\npp}, this
distinction does not make much of a difference. For
$\alpha=o(n^{1/4})$, the original and the modified conjectures are
equivalent, and the original REM conjecture holds, while for $\alpha$
growing like a constant times $n^{1/4}$, both the original and the
modified REM conjectures fail. So for the {\npp}, the threshold for
the validity of the REM conjecture is $n^{1/4}$, not $\sqrt n$ as
one might have naively guessed.

\subsection{The SK Spin Glass}

In a follow-up paper to \cite{rem1},
Bauke and Mertens generalized the local REM conjecture
for the {\npp} to a large class of disordered systems,
including
 many  other models
of combinatorial optimization as well as several
spin glass models \cite{rem2}.
Motivated by this conjecture,
Bovier and Kurkova developed an axiomatic approach to
Poisson convergence, covering, in particular
many types
of spin glasses like the Edwards-Anderson model and the
Sherrington-Kirkpatrick model.

The Sherrington-Kirkpatrick  model (SK model)
is  defined by the energy function
\begin{equation}
\label{E-SK}
E(\bss)=\frac 1{\sqrt n}\sum_{i,j=1}^nX_{ij}\sigma_i\sigma_j.
\end{equation}
As before, $\bss$ is a spin configuration in $\{-1,+1\}^n$,
but now the random input is given in terms of $n^2$ random
variables $X_{ij}$ with $i,j\in\{1,\dots,n\}$,
usually taken to be i.i.d. standard normals.
Again $E(\bss)=E(-\bss)$, leading to $N=2^{n-1}$ {\em a priori}
different energies $E_1\leq E_2\leq\dots\leq E_N$.
Note, however, that the normalization in
\eqref{E-SK} corresponds
to typical energies of order $\sqrt n$
and maximal energies of order $n$, in accordance
with the standard physics notation.

Consider an energy scale $\alpha_n\geq 0$, and let
$E_{r+1}$ be the first energy above $\alpha_n$.
To obtain the REM approximation for the
SK model, we observe that the
random variable $E(\bss)$ is a
Gaussian with
density
\begin{equation}
\label{g-SK}
\tilde g(x)=\frac 1{\sqrt {2\pi n}}e^{-x^2/2n}.
\end{equation}
The REM approximation
for the SK model therefore suggests that
 the rescaled energy
spectrum
$(E_{r+1}-\alpha_n)\tilde\xi_n^{-1}$, $(E_{r+2}-\alpha_n)\tilde\xi_n^{-1}$,
$\dots$
with rescaling factor
\begin{equation}
\label{tilde-xi-n}
\tilde\xi_n=\sqrt{{2\pi n }}\, 2^{-(n-1)}
e^{\frac{\alpha_n^2}{2n}}
\end{equation}
converges to a Poisson process with density one.
Recalling that typical energies are now  of
order $\sqrt n$, the local
REM conjecture for {\em typical energies} in the
SK model thus claims that for $\alpha_n=O(\sqrt n)$,
the rescaled energy spectrum converges to a Poisson
process with density one, with overlaps which
again tend to zero as $n\to\infty$.

This conjecture was proved in a very nice paper
by Bovier and Kurkova \cite{bovier:kurkova:05}.  In fact, they
proved that the conjecture remains
valid as long as $\alpha_n=O(n^\eta)$ with
 $\eta<1$.  To get some insight into still faster
growing $\alpha_n$, Bovier and Kurkova then
considered the generalized
random energy model (GREM) of Derrida  \cite{derrida:85}.
For this model, they proved \cite{bovier:kurkova:05b}
that the local REM conjecture
holds up to $\alpha_n=\beta_0 n$, where $\beta_0$ is the
inverse transition temperature of the GREM, and
fails once $\alpha_n$ exceeds this threshold.
Based on these results for the GREM,
Bovier and Kurkova then suggested
\cite{bovier:kurkova:05c}
that
the $\beta_0$ might be
the threshold for other
disordered
spin systems as will.

As we will show in this paper,
this is not the case, at least not for the SK model,
for which we prove that the REM conjecture holds up
to the threshold $\alpha_n=o(n)$,
and becomes invalid
as soon as $\limsup \alpha_n/n>0$, see Theorem \ref{thm:SK}
below for the precise statement.
Thus even the scaling with $n$ of the threshold
does not obey the naive expectation derived from the
GREM.
Note that for the SK model
there is no difference between the original
REM conjecture and the modified REM conjecture, since the density
of $E(\bss)$ is Gaussian for all energy scales.

\subsection{Organization of the Paper}

This paper is organized as follows. In the next section, we
precisely state the assumptions on our model and formulate
our main results, see Theorems~\ref{thm:growing} and
\ref{thm:SK}.  In Section~\ref{sec:Proof-Strat}, we then
describe our main proof strategy for the {\npp}.  Since
the proof strategy for the SK model only requires minor
modifications (the proof is, in fact, much easier), we
defer the discussion of this model to the last subsection,
Section~\ref{sec:SK-Strat}.  The next four sections contain
the details of the proof: As a warmup, we start with the
{\npp} with Gaussian noise, where our strategy is most
straightforward.  Next, in Section~\ref{sec:Poisson-gen},
we move to the {\npp} with a general distribution.  This
section contains the meat of our proof: the establishment of
a large deviations estimate for the probability density
of several (weakly dependent) random variables.  In
Section~\ref{sec:SK-Proof}
we give the proof of Theorem~\ref{thm:SK}, and
in Section~\ref{sec:Aux} we establish several auxiliary
results needed in the rest of the paper.  We conclude
the paper with a section summarizing our results and
discussing possible extensions, Section~\ref{sec:Discuss}.

\section{Statement of Results}

\subsection{Number Partitioning}
\label{sec:npp-results}

Let $X_1,...,X_n$ be independent random variables distributed
according to the common density function $\rho(x)$.
We assume that $\rho$ has second moment
one
and satisfies the bound
\eq
\label{dist}
\int_{-\infty}^{\infty} \abs{\rho(x)}^{1+\epsilon} \;
dx < \infty
\en
for some $\epsilon > 0$. Since neither the distribution of the
overlaps nor the energy spectrum changes if we replace $\rho(x)$ by
$\frac 12(\rho(x)+\rho(-x))$, we will assume that
$\rho(x)=\rho(-x)$.   We use the symbol $\PR_n(\cdot)$ to denote the
probability with respect to the joint probability distribution of
$X_1,...,X_n$, and the symbol $\EX_n(\cdot)$ to denote expectations
with respect to $\PR_n(\cdot)$.

As in the introduction, we represent the $2^{n}$ partitions of the
integers $\{1,..,n\}$ as spin configurations $\bss \in
\{-1,+1\}^n$
and
define the energy of $\bss$ as in \eqref{ene}.
Recalling that the distribution of $E(\bss)$ does not
depend on $\bss$, let $g_n(\cdot)$ be the density
of $E(\bss)$,
and let $\xi_n$ be the modified scaling factor
\begin{equation}
\label{xi-n-mod}
\xi_n=\frac 1{2^{n-1}g_n(\alpha_n)}.
\end{equation}
We now introduce a continuous time process $\{N_n(t)\colon t\geq
0\}$ where $N_n(t)$ is defined
 as the number of points in the energy spectrum
that lie
between $\alpha_n$ and $\alpha_n+t\xi_n$.

Let
$E_1,\ldots ,E_N$
be
the increasing spectrum of the energy values
corresponding to the $N = 2^{n-1}$ distinct partitions.
Given $\alpha_n\geq 0$,
let $r_n$ be the random variable
defined by $E_{r_n}<\alpha_n\leq E_{r_n+1}$.
For $j>i>0$, we then define the rescaled overlap
$Q_{ij}$  as the random  variable
\begin{equation}
\label{Q-def} Q_{ij}=
\frac {1}4
\sum_{\bss,\tilde\bss}
n^{1/2}q(\bss,\tilde\bss)
\end{equation}
where the sum goes over the four pairs of configurations with
$E(\bss)=E_{r_n+i}$ and $E(\tilde\bss)=E_{r_n+j}$. Instead of the
overlap $Q_{ij}$, we will sometimes consider the following variant:
consider two distinct configurations $\bss$ and $\tilde\bss$ chosen
uniformly at random from all pairs of distinct configurations.  We
then define $Q_{n,t}$ as the rescaled overlap
$n^{1/2}q(\bss,\tilde\bss)$ conditioned on the event that $E(\bss)$
and $E(\tilde\bss)$ fall into the energy interval
$[\alpha_n,\alpha_n+t\xi_n]$. We will refer to $Q_{n,t}$ as the {\em
overlap between two typical configurations contributing to}
$N_n(t)$.

The main results of this paper are statements ii) and iii) of the
following theorem.  The first statement is a corollary of the proof
of ii) and iii) and implies that for $\alpha_n=o(n^{1/4})$, the
original and the modified REM conjecture are equivalent.

\begin{theorem}
\label{thm:growing}
Let $X_1,...,X_n\in\mathbb R$ be i.i.d.~random variables
drawn from a probability distribution with
second moment one and even density $\rho$.
If $\rho$ obeys the assumption \eqref{dist}
for some $\epsilon>0$
and has
a Fourier transform that is analytic in a neighborhood
of zero,
then the following holds:

i) Let $g_n(\cdot)$ be the density of  $E(\bss)$,
and let $g(\alpha)=\sqrt{{2}/{\pi}}e^{-\alpha^2/2}$.
If $\alpha_n=o(n^{1/4})$, then
$g_n(\alpha_n)=g(\alpha_n)(1+o(1))$.

ii) Let $\alpha_n=o(n^{1/4})$, and
let $j>i>0$ be arbitrary integers not depending on $n$.
As $n\to\infty$, the process $N_n(t)$ converges weakly to a
Poisson process with density one,
and both $Q_{ij}$ and $Q_{n,t}$ converge
in distribution
to standard normals.

iii) Let $\alpha_n=\kappa n^{1/4}$ for some finite
$\kappa>0$.  Then $\EX_n[N_n(t)]=t+o(1)$,
but the process $N_n(t)$ does not
converge to a Poisson process,
and $Q_{n,t}$ does not converge to a standard normal.
\end{theorem}


In order to prove the above  theorem, we will analyze the
factorial moments of the process $Z_n(t)$.  We will show that for
$\alpha_n=o(n^{1/4})$, they converge to those of a Poisson process,
and for $\alpha_n=\kappa n^{1/4}$ with $\kappa>0$, they do not
converge to the moments of a Poisson process.  Together with
suitable upper bounds on the moments of $Z_n(t)$, this allows us to
establish non-convergence to Poisson for $\kappa>0$, but,
unfortunately, it does not allow us to establish convergence to some
other distribution.

The situation is slightly more ``constructive'' for the overlap
distribution: here we are able to determine the limiting
distribution of $Q_{n,t}$ for $\alpha_n=\kappa n^{1/4}$ with
$\kappa>0$.  In this regime, the distribution of $Q_{n,t}$ converges
to a convex combination of two shifted Gaussians: with probability
$1/2$ a Gaussian with mean $\kappa^2$ and variance one, and with
probability $1/2$ a Gaussian with mean $-\kappa^2$ and variance one,
so in particular $Q_{n,t}$ is not asymptotically normal. As we will
see, this is closely connected to the failure of Poisson
convergence; see Remark~\ref{rem:overlap} in Section~\ref{sec:Gauss}
and Remark~\ref{rem:overlap-non-gauss} in
Section~\ref{sec:Poisson-gen} below.

\subsection{SK Spin Glass}
We consider the SK model with energies given
by \eqref{E-SK} and random coupling $X_{ij}$
which are i.i.d.~standard normals. Let
$N=2^{n-1}$, and let  $E_1,\dots,E_N$ and
 $\bss^{(1)},\dots,\bss^{(N)}$ be as defined in the
 introduction.

Given an energy scale $\alpha_n\geq 0$
and two integers $j>i>0$,
we  again introduce
$Q_{ij}$ as the  random  variable   defined in \eqref{Q-def},
with
$r_n$ given by the condition that $E_{r_n}<\alpha_n\leq E_{r_n+1}$.
Finally, we  define
$N_n(t)$ to be the number of points in the energy spectrum
of \eqref{E-SK}
that lie
between $\alpha_n$ and $\alpha_n+t\tilde\xi_n$,
with $\tilde\xi_n$ given by \eqref{tilde-xi-n}.
We say that the local REM conjecture holds if
$N_n(t)$ converges weakly to a Poisson
process with density one, and $Q_{ij}$ converges in distribution
to a standard normal for all $j>i>0$.

Our proofs for the {\npp} can then be easily generalized to
give the following theorem.

\begin{theorem}
\label{thm:SK}
There exists a constant $\epsilon_0>0$ such the
following statements hold for all sequences
of positive real numbers $\alpha_n$ with $\alpha_n\leq\epsilon_0n$:

(i) $\EX(N_n(t))\to t$ as $n\to\infty$.

(ii) The local REM conjecture for the SK model holds
if and only if $\alpha_n=o(n)$.
\end{theorem}

\section{Proof Strategy}
\label{sec:Proof-Strat}

In this section we describe our proof strategy for
Theorems~\ref{thm:growing} and \ref{thm:SK}.  We explain our ideas
using the example of the {\npp}, referring to the SK model
only in the last subsection, Section~\ref{sec:SK-Strat}.

\subsection{Factorial moments}
\label{sec:Fact-Moment}

Consider a finite family of non-overlapping intervals $[c_1,d_1]$,
$\dots$, $[c_m,d_m]$ with $d_\ell> c_\ell\geq 0$, and let
$\gamma_\ell=d_\ell-c_\ell$. Weak convergence of the process
$\{N_n(t)\colon t\geq 0\}$ to a Poisson process of density
one is equivalent to the statement that for each such family, the
increments $N_n(d_1)-N_n(c_1)$, $\dots$, $N_n(d_m)-N_n(c_m)$
converge to independent Poisson random variables with rates
$\gamma_1,\dots,\gamma_m$.

Let $Z_n(a,b)$ be the number of point in the energy spectrum that
fall into the interval $[a,b]$.  We then rewrite
$N_n(d_\ell)-N_n(c_\ell)$ as $Z_n(a_n^\ell,b_n^\ell)$, where
$a_n^\ell=\alpha + c_\ell \xi_n$ and $b_n^\ell=\alpha + d_\ell
\xi_n$, with $\xi_n$ defined in equation  \eqref{xi-n-mod}. We
prove convergence of the increments to independent Poisson random
variables by proving convergence of the multidimensional factorial
moments, i.e., by proving the following theorem.  To simplify
our notation, we will henceforth drop the index $n$ on both
the symbol $\EX_n$ and $\PR_n$.

\begin{theorem}
\label{thm:main1} Let $\rho$ be as in  Theorem \ref{thm:growing},
let $\alpha_n=o(n^{1/4})$ be a sequence of positive real numbers,
let $m$ be a positive integer, let $[c_1,d_1]$, $\dots$, $[c_m,d_m]$
be a family of non-overlapping, non-empty intervals, and let
$(k_1,..,k_m)$ be an arbitrary $m$-tuple of positive integers. For
$\ell=1,\dots,m$, set $a_n^\ell= \alpha_n + c_\ell\xi_n$, $b_n^\ell=
\alpha_n + d_\ell\xi_n$, and $\gamma_\ell=d_\ell-c_\ell$. Under
these conditions, we have \eq \label{mulfact} \lim_{n\ra\infty}
\mathbb{E}[\prod_{\ell=1}^m (Z_n(a_n^{\ell},b_n^{\ell}))_{k_{\ell}}]
= \prod_{\ell=1}^m \gamma_\ell^{k_{\ell}}, \en
 where, as usual, $(Z)_k=Z(Z-1)\dots (Z-k+1)$.
\end{theorem}

Theorem~\ref{thm:main1} establishes that  $\{N_n(t)\colon t\geq 0\}$
converges to a Poisson
process with density one
if $\alpha_n=o(n^{1/4})$.  The convergence of the
overlaps is an easy corollary to the proof of this theorem.
The details are exactly the same as
in \cite{part1},
and will not be repeated here.

The failure of
Poisson convergence  for faster growing
$\alpha_n$
is easiest to explain for the case in which
$X_1,\dots,X_n$ are standard normals, since this
does not require us to distinguish between the original
and the modified REM conjectures.
Our proof is again based on the
analysis of
the factorial moments of
$N_n(t)$.

More precisely, we will show that  $E[N_n(t)]$ converges
to $t$, while the second factorial moment, $E[(N_n(t))_2]$ does
not converge to $t^2$.  Note that this fact by itself is not
enough to exclude convergence to a Poisson process since
convergence of the factorial moments is, in general, only a
sufficient condition for weak convergence to a Poisson process.
But combined with suitable estimates on the growth of the third
moment, the fact that $\EX[(N_n(t))_2]$ does not converge to $t^2$
is enough. This
follows from the following lemma, which is an easy
consequence of a standard theorem on uniformly integrable sequences
of random variables
(see, e.g., Theorem 25.12 and its
corollary in \cite{billingsley}).

\begin{lemma}
\label{lem:non-poisson} Let
$Z_n\geq 0$ be a sequence of random
variables such that $\EX[Z_n^{r}]$
is bounded uniformly in $n$
for some $r<\infty$.
If $Z_n$ converges weakly to a Poisson random
variable with rate
$\gamma>0$, then
$\lim_{n\to\infty}\EX[(Z_n)_k]=\gamma^k$
for all $k<r$.
\end{lemma}

Combined with this lemma, the next theorem establishes
the third statement of Theorem~\ref{thm:growing}
if the weights $X_1,\dots,X_n$ are Gaussian.
The proof for non-Gaussian weights will be given
in Section~\ref{sec:Fast-Grow}.

\begin{theorem}
\label{thm:main2}
Let $X_1,...,X_n\in\mathbb R$ be i.i.d.~random variables
with normal distribution, and let $\alpha_n=o(\sqrt n)$.

(i) $\EX[N_n(t)]=t+o(1)$ for all fixed $t>0$.

(ii) Let $m$, $a_n^\ell$, $b_n^\ell$,
$\gamma_\ell$, and $k_1,..,k_m$ be
as in Theorem~\ref{thm:main1}.  For
$k=\sum_{\ell=1}^m k_\ell\geq 2$, we then have
\eq
\label{mulfact-gauss}
\mathbb{E}[\prod_{\ell=1}^m (Z_n(a_n^{\ell},b_n^{\ell}))_{k_{\ell}}] =
\Bigl(\prod_{\ell=1}^m
\gamma_\ell^{k_{\ell}}\Bigr)
e^{\frac{k(k-1)}{4n}\alpha_n^4}
e^{O(\alpha_n^6n^{-2})+o(1)}.
\en
\end{theorem}

Theorem~\ref{thm:main2} will be proved in Section~\ref{sec:Gauss},
 and Theorem~\ref{thm:main1} will be proved in
Section~\ref{sec:Poisson-gen}.
In the remainder of this section,
we will map out the general proof strategy, and in the
process, establish several properties
which will be used in the proofs of Theorems
\ref{thm:main1} and \ref{thm:main2}.

\subsection{Analysis of first moment}
\label{sec:First-Moment-Heur}

In order to analyze the first moment
we rewrite $Z_n(a_n^{\ell},b_n^{\ell})$
as
\eq
\label{Z1} Z_n(a_n^{\ell},b_n^{\ell}) =
 \sum_{\bss} I^{(\ell)}(\bss)
\en
where $I^{(\ell)}(\bss)$ is $1/2$ times
an indicator function of the event that the energy $E(\bss)$
falls into the interval $[a_n^{\ell},b_n^{\ell}]$ (the factor $1/2$
compensates for the fact that the sum in
\eqref{Z1} goes over all configurations $\bss\in\{-1,+1\}^n$,
and therefore counts every distinct partition twice).
Taking expectations of both sides we see that
\eq
\label{Exp-Z1}
\EX[Z_n(a_n^{\ell},b_n^{\ell})] =
\frac 12 \sum_{\bss} \PR\Bigl(E(\bss)\in
[a_n^{\ell},b_n^{\ell}]\Bigr).
\en
Since
we have taken $\rho (x) = \rho (-x)$,
the distribution of $E(\bss)$ is identical for all $\bss$,
so that the probability on the right hand side
is independent of $\bss$.
In order to prove that
$\EX[Z_n(a_n^{\ell},b_n^{\ell})]=\gamma_\ell(1+o(1))$,
we will therefore want to prove that
\begin{equation}
\label{First-moment-prob}
\PR\Bigl(E(\bss)\in [a_n^{\ell},b_n^{\ell}]\Bigr)=
2^{-(n-1)}\gamma_\ell(1+o(1)).
\end{equation}
Rewriting the left hand side as
\begin{equation}
\label{Exp-Z1-dens}
\PR\Bigl(E(\bss)\in [a_n^{\ell},b_n^{\ell}]\Bigr)
=\int_{a_n^{\ell}}^{b_n^{\ell}}
g_n(y)dy,
\end{equation}
where $g_n(\cdot)$ is the density of $E(\bss)$ with respect to the
Lesbesgue measure on $\R_+=\{x\in\R\colon x\geq 0\}$, and recalling
the definition $\xi_n = [2^{n-1}g_n(\alpha_n)]^{-1}$,
we see that the bound \eqref{First-moment-prob} is equivalent to the
statement that $g_n(y)=g_n(\alpha_n)(1+o(1))$ whenever
$y=\alpha_n+O(\xi_n)$.

This bound is easily established in the Gaussian case, where it holds
as long as $\alpha_n\leq c\sqrt n$ for some $c<\sqrt{2\log 2}$.
Indeed, let us write $E(\bss)$ as $|H(\bss)|$, where
\begin{equation}
\label{H-def}
H(\bss) = \frac{1}{\sqrt{n}}\sum_{i=1}^n \sigma_i X_i.
\end{equation}
If the random variables $X_1,\dots,X_n$ are standard Gaussians,
the random variables $H(\bss)$ are standard Gaussians as well,
implying that $E(\bss)$ has density
\begin{equation}
\label{g.Gauss}
g(y)=\sqrt{\frac{2}{\pi}}e^{- y^2/2}
\end{equation}
with respect to the Lesbesgue measure on $\R_+=\{x\in\R\colon x\geq
0\}$. If $\alpha_n\leq c\sqrt n$ for some $c<\sqrt{2\log 2}$,
then $\xi_n=o(e^{-\epsilon
n})$ for some $\epsilon > 0$, implying that for
$y=\alpha_n+O(\xi_n)$, we have $g(y)= g(\alpha_n)(1+O(\sqrt n\xi_n))
=g(\alpha_n)(1+o(1))$, as desired. This proves
\eqref{First-moment-prob} and hence the bound
$\EX[Z_n(a_n^{\ell},b_n^{\ell})]=\gamma_\ell(1+o(1))$ provided
$\alpha_n\leq c\sqrt n$ for some $c<\sqrt{2\log 2}$. Note that this
already establishes the first moment bound stated in
Theorem~\ref{thm:main2}.

For more general distributions,
the proof is more complicated since
$g_n(\alpha_n)$ is no longer given by a simple formula
like \eqref{g.Gauss}. But given our assumptions
on $\rho$, we will be able to show that
under the assumption that $\alpha_n=o(\sqrt n)$, it can
be written in
the form
\begin{equation}
\label{gn.largedev}
g_n(\alpha_n)=\sqrt{\frac 2\pi}e^{-nG(\alpha_nn^{-1/2})}(1+o(1))
\end{equation}
where $G$ is an even function which is analytic in a neighborhood
of zero and satisfies the bound
\begin{equation}
\label{G.bound}
G(x)=\frac {x^2}2+O(x^4).
\end{equation}
The proof of the representation \eqref{gn.largedev} involves
an integral representation for $g_n$ combined with a
steepest descent analysis and will be given in
Section~\ref{section:firstmoment}.

The bounds \eqref{gn.largedev} and \eqref{G.bound} have several
immediate consequences. First, they clearly imply that
$g_n(\alpha_n)=g(\alpha_n)(1+o(1))$ if $\alpha_n=o(n^{1/4})$,
proving the first statement of Theorem~\ref{thm:growing}.  Second,
they imply that $\xi_n$ decays exponentially in $n$ if
$\alpha_n=o(\sqrt n)$. Using the bounds \eqref{gn.largedev} and
\eqref{G.bound} once more, we conclude that
$g_n(y)=g_n(\alpha_n)(1+o(1)+O(\alpha_n\xi_n))=
g_n(\alpha_n)(1+o(1))$ whenever $y=\alpha_n+O(\xi_n)$. For
$\alpha_n=o(\sqrt n)$, we therefore get convergence of the first
moment, $\EX[Z_n(a_n^{\ell},b_n^{\ell})]=\gamma_\ell(1+o(1))$,
implying in particular that $\EX[Z_n(t)]=t+o(1)$ as claimed in the
last statement of Theorem~\ref{thm:growing}.

\subsection{Factorial moments
as sums over pairwise distinct configurations}

Next we turn to higher factorial moments.
Before studying these factorial moments, let us consider
the standard  moments
$\mathbb{E}
[\prod_{\ell=1}^m (Z_n(a_n^{\ell},b_n^{\ell}))^{k_{\ell}}]$.
In view of \eqref{Z1}, these moments can  be written as a
sum over $k$ configurations $\bss^{(1)},\dots, \bss^{(k)}$,
where $k=\sum_{\ell=1}^m k_{\ell}$.
As already observed in \cite{borgs:chayes:pittel:01},
the factorial moments can be expressed in a similar way,
the only difference being that the sum over configurations
is now a sum over pairwise distinct configurations, i.e.,
configurations $\bss^{(1)},\dots, \bss^{(k)}\in\{-1,+1\}^n$
such that $\bss^{(i)}\neq\pm\bss^{(j)}$
for all $i\neq j$.  Explicitly,
\begin{equation}
\begin{aligned}
\label{Zm-multi}
\mathbb{E}
\Bigl[\prod_{\ell=1}^m
&(Z_n(a_n^{\ell},b_n^{\ell}))_{k_{\ell}}\Bigr]
=\sum_{\bes\bss^{(1)} ,\dots, \bss^{(k)}:
\\ \bss^{(i)}\neq\pm\bss^{(j)}\es}
  \mathbb E\Bigl[\prod_{j=1}^{k} I^{(\ell(j))}(\bss^{(j)})\Bigr]
\\
&=\frac 1{2^{k}}
\sum_{\bes\bss^{(1)} ,\dots, \bss^{(k)}:
\\ \bss^{(i)}\neq\pm\bss^{(j)}\es}
\PR\Bigl(E(\bss^{(j)})\in [a_n^{\ell(j)},b_n^{\ell(j)}]
  \text{ for all } j=1,\dots,k\Bigr),
\end{aligned}
\end{equation}
where the sums
run over pairwise distinct configurations
and
$\ell(j)=1$ if $j=1,\dots,k_1$, $\ell(j)=2$ if $j=k_1+1,\dots, k_1+k_2$,
and so on.  See \cite{part1} for the
(straightforward) derivation of \eqref{Zm-multi}.

In order to prove
convergence of the higher factorial moments, we therefore would like
to show that the probability
$$
\PR\left(E(\bss^{(j)})\in [a_n^{\ell(j)},b_n^{\ell(j)}]\text{ for all j}\right)
$$
asymptotically factors into the product
$\prod_j\PR(E(\bss^{(j)})\in [a_n^{\ell(j)},b_n^{\ell(j)}])$.
Unfortunately, this asymptotic factorization does not hold
for arbitrary families of distinct configurations
$\bss^{(1)}\dots \bss^{(k)}$.  This problem is already present
for $\alpha_n$ that are bounded as $n\to\infty$ (see \cite{part1}),
and -- in a milder form -- it is even present for the special
case of $\alpha_n=0$ treated in \cite{borgs:chayes:pittel:01}.

\subsection{Typical and atypical configurations}

As in \cite{part1} and \cite{borgs:chayes:pittel:01},
it is useful to distinguish between ``typical'' and  ``atypical''
 sets of configurations
$\bss^{(1)},...,\bss^{({k})}$ when analyzing the right hand
side of \eqref{Zm-multi}.
To this end, we consider the
matrix $M$
formed by the row vectors $\bss^{(1)},\dots,\bss^{(k)}$.
Given a vector
$\bsd \in \{-1,1\}^k$, we then define
$n_\bsd(\bss^{(1)},...,\bss^{(k)})$ as the
number of times the column vector $\bsd$ appears in the matrix $M$,
\begin{equation}
\label{n-delta}
n_\bsd = n_\bsd(\bss^{(1)},...,\bss^{(k)}) = |
\{j\leq n: (\sigma_j^{(1)},...,\sigma_j^{(k)}) = \bsd\}|.
\end{equation}
If
$\bss^{(1)},\dots,\bss^{(k)}\in\{-1,+1\}^n$ are chosen
independently and uniformly at
random, then
the expectation of $n_\bsd$ is  equal to $n2^{-k}$
for all $\delta\in\{-1,+1\}^k$.  By a standard martingale argument,
for most configurations,
the difference  between $n_\bsd$ and $n2^{-k}$ is therefore not
much larger than $\sqrt n$.  More precisely,
for any $\lambda_n\to\infty$ as $n\to\infty$, all but a vanishing
fraction of the
configurations
$\bss^{(1)},\dots,\bss^{(k)}$ obey the condition
\eq
\label{const-lan}
\max_{\bsd} | n_{\bsd}(\bss^{(1)},...,\bss^{(k)}) -
\frac{n}{2^k}| \leq \sqrt{n} \,\lambda_nn ,
\en
see
Lemma 3.9 in \cite{part1} for a proof.

The proof of Theorems~\ref{thm:main1} and \ref{thm:main2} now
proceeds in two steps.  First, we show that the contribution
of the configurations that violate
\eqref{const-lan}
 is negligible as $n\to\infty$,
and second we analyze the configurations satisfying
\eqref{const-lan}.  It turns out that first part is quite
complicated and requires distinguishing several sub-classes of
configurations, but  this analysis has already been carried out in
\cite{part1}, resulting in bounds that are sharp enough for
growing $\alpha_n$ as well.  So the only additional work
needed is a sharp analysis of the typical configurations.

The  next lemma
summarizes the main results from \cite{part1} needed in
this paper.  To state it, we define $R_{n,k}(\lambda_n)$ as
\begin{equation}
\label{Error-def}
R_{n,k}(\lambda_n)=
 \frac 1{2^k}
\sump_{\bss^{(1)},\cdots ,\bss^{(k)}}
    \PR\Bigl(E(\bss^{(j)})\in [a_n^{\ell(j)},b_n^{\ell(j)}]
     \text{ for all } j=1,\dots,k\Bigr),
\end{equation}
where the sum runs over pairwise distinct configurations
$\bss^{(1)},\dots, \bss^{(k)}$ that are either linearly
dependent or
violate the bound \eqref{const-lan}.
We
also use the notation
\begin{equation}
q_{\max}=\max_{i\neq j}|q(\bss^{(i)},\bss^{(j)})|
\end{equation}
for the maximal off-diagonal overlap of
$\bss^{(1)},\dots, \bss^{(k)}$.

\begin{lemma}
\label{lem:comb-bds} Let $\lambda_n$ be a sequence of positive
numbers.

\noindent
\begin{enumerate}

\item  Then
the number of configurations
$\bss^{(1)},...,\bss^{(k)}$ that violate the condition
\eqref{const-lan}
is bounded by
$2^{nk}2^{k+1}e^{-\frac 12\lambda_n^2}$.

\item
   Assume that both $\alpha_n$ and $\lambda_n$ are of order
$o(\sqrt n)$. Then there are constants $c,C<\infty$ depending only
on $k$, $\gamma_1,\dots,\gamma_k$, and the sequence $\lambda_n$,
such that for $n$ sufficiently large we have
\begin{equation}
R_{n,k}(\lambda_n) \leq Cn^{c}
e^{\frac 12 k\alpha_n^2(1+o(1))}
\bigl(\xi_n^{1/n_0} + e^{-\lambda_n^2/2}\bigr).
\end{equation}
Here
$n_0=(1+\epsilon)/\epsilon$ with $\epsilon$ as in \eqref{dist}.

\item Let
$\bss^{(1)},...,\bss^{(k)}$ be
an arbitrary set of row vectors
satisfying \eqref{const-lan}.  Then
\begin{equation}
\label{olap}
q_{\max}
\leq
2^k\frac{\lambda_n}{\sqrt n},
\end{equation}
and hence for $\lambda_n=o(\sqrt n)$ and
$n$ sufficiently large, we have that
$\bss^{(1)},...,\bss^{(k)}$
are linearly independent.
\end{enumerate}
\end{lemma}

\begin{proof}
Statements (1) and (3) are copied from Lemma 3.8 in \cite{part1},
while statement (2)  is a consequence of the bounds (3.65) and
(3.67) in \cite{part1} and the fact that
$2^n\xi_n=e^{\frac 12\alpha_n^2(1+o(1))}e^{O(1)}$, a bound which
follows immediately from \eqref{gn.largedev}, \eqref{G.bound}, and
the definition \eqref{xi-n-mod} of $\xi_n$.
\end{proof}

\subsection{Factorization for typical configurations}
\label{sec:Typ-Fact-Heur}

In view of Lemma~\ref{lem:comb-bds}, it will be enough to analyze
the expectations on the right hand side
of \eqref{Zm-multi} for configurations
$\bss^{(1)},...,\bss^{(k)}$ that satisfy \eqref{const-lan}
and are linearly independent, provided we choose $\lambda_n$ in
such a way that $\alpha_n=o(\lambda_n)$,
$\lambda_n=o(\sqrt n)$ and $e^{-\lambda_n^2/2}$
decays faster than any power of $n$.

Consider therefore a family of
linearly independent configurations satisfying the
condition \eqref{const-lan}.
  The main technical result of
this paper is   that  the following approximate factorization
statement
\begin{equation}
\begin{aligned}
\label{Approx-Fact}
\PR\Bigl(E(\bss^{(j)})&\in [a_n^{\ell(j)},b_n^{\ell(j)}]
  \text{ for all } j=1,\dots,k\Bigr)
  \\
&=\prod_{j=1}^k \PR\Bigl(E(\bss^{(j)})\in
[a_n^{\ell(j)},b_n^{\ell(j)}]
 \Bigr)
e^{O(\alpha_n^2 q_{\text{max}})+o(1)}
\end{aligned}
\end{equation}
is valid whenever $n$ is large enough and
$\bss^{(1)},...,\bss^{(k)}$ obey the condition \eqref{const-lan}.
(For the non-Gaussian case, we will also need that
$\alpha_n=o(n^{1/4})$,  while the assumption
$\alpha_n=o(\sqrt n)$ stated at the beginning of this subsection is
enough in the Gaussian case.)

If $X_1,\dots,X_n$ are standard normals,
the bound \eqref{Approx-Fact}
is quite intuitive and not hard to prove.  Indeed,
let $H(\bss)$ be the random variable
defined in \eqref{H-def}, and let
$\kappa^{(k)}(\cdot)$
be the joint density of
$H(\bss^{(1)})$, $\dots$, $H(\bss^{(k)})$.
Using the notation $\bold x$ and $\bold y$ for vectors
with components $x_1,\dots,x_k$ and $y_1,\dots,y_k$,
respectively,
we then express the joint density
of $E(\bss^{(1)})$, $\dots$, $E(\bss^{(k)})$
with respect to the Lebesque measure on $\R_+^k$
as
\begin{equation}
\label{H-Gauss-density-k}
 g^{(k)}(\bold y)=
\sum_{\bes x_1,\dots x_k:\\ y_i=\pm x_i\es}
\kappa^{(k)}(\bold x).
\end{equation}
 If
$X_1,\dots,X_n$ are standard Gaussians, the joint distribution
of $H(\bss^{(1)})$, $\dots$,
$H(\bss^{(k)})$ is Gaussian as well, with mean zero and
covariance matrix
\begin{equation}
\label{C-Gauss}
C_{ij}
=E[H(\bss^{(i)})H(\bss^{(j)})]=
q(\bss^{(i)},\bss^{(j)}),
\end{equation}
leading to the representation
\begin{equation}
\label{Gauss-density-k}
\kappa^{(k)}(\bold x)=
\frac 1{(2\pi)^{k/2}}\frac 1{\sqrt{\det C}}
\exp\Bigl(-\frac 12(\bold x, C^{-1}\bold x)\Bigr),
\end{equation}
whenever $C$ is invertible.  As usual, $C^{-1}$ denotes
the matrix inverse of $C$, and
$(\bold x, C^{-1}\bold x)=\sum_{i,j}x_i C^{-1}_{ij}x_j$.
Observing that  $C^{-1}_{ij}=\delta_{ij}+O({q_\text{max}})$,
this
 immediately leads to the bound
\eqref{Approx-Fact};  see Section~\ref{sec:Gauss} for details.

In the non-Gaussian case, we do not have an explicit
formula for the joint density of $E(\bss^{(1)})$, $\dots$,
$E(\bss^{(k)})$, thus making the proof of
the approximate factorization formula \eqref{Approx-Fact}
much more difficult.  Basically, the proof
requires a multi-dimensional
local limit theorem  for a case in which the arguments
of the probability density under investigation grow with
$n$.  In order to prove this local limit theorem,
we will use an integral representation for the
probabilities on the left hand side of
\eqref{Approx-Fact} and then use a saddle point
analysis to prove \eqref{Approx-Fact}.
In contrast to \cite{part1} and
\cite{borgs:chayes:pittel:01},
where it was sufficient to analyze the saddle point of
the integral in the original domain of integration, the
case with growing $\alpha_n$ considered here requires
a more sophisticated
analysis, involving the shift to a complex saddle point in
a complex space of dimension $k$; see
Section~\ref{sec:Poisson-gen} for the details.

Let us close this section by showing how to complete the proof of
Theorem~\ref{thm:main1} once the bound \eqref{Approx-Fact} is
established. To this end, let us first consider the case where
$\alpha_n$ grows somewhat more slowly  than $o(n^{1/4})$, in
particular assume that  $\alpha_n=o(n^{1/6})$.  Choosing
$\lambda_n=n^{1/6}$, we then invoke the bound \eqref{olap} from
Lemma~\ref{lem:comb-bds} (3) to conclude that
$q_{\text{max}}=O(n^{-1/3})$ and $\alpha_n^2q_{\text{max}}=o(1)$
whenever $\bss^{(1)},...,\bss^{(k)}$ satisfy the bound
\eqref{const-lan}. For a family of configurations satisfying
\eqref{const-lan}, the the multiplicative error term in the joint
probability on the right hand side of \eqref{Approx-Fact} is
therefore equal to $1+o(1)$, leading to asymptotic factorization of
the probabilities on the right hand side of \eqref{Zm-multi}. Using
Lemma~\ref{lem:comb-bds} (2) to bound the sum over families of
configurations not satisfying \eqref{const-lan}, and
Lemma~\ref{lem:comb-bds} (1) to show that the number of families of
configurations satisfying \eqref{const-lan} is equal to
$2^{nk}(1+o(1))$, this gives the bound \eqref{mulfact} from
Theorem~\ref{thm:main1}.

For still more quickly growing $\alpha_n$, it is not enough
to use the worst case bound \eqref{olap}.  Instead, we
would like to use that a typical family of configurations
has maximal off-diagonal overlap
of order $n^{-1/2}$.  For
a typical family of configurations, we therefore
get asymptotic factorization as long as
$\alpha_n=o(n^{1/4})$.
The next lemma, to be proved in Section~\ref{sec:Aux},
implies that  the error term coming
from atypical configurations does not destroy the
asymptotic factorization.

\begin{lemma}
\label{lem:sum-factor}
Let ${c}>0$, let $k$ be a positive integer,
and let
$\alpha_n$ be a
sequence of positive numbers
such that
$\alpha_n n^{-1/4}\to 0$ as $n\to\infty$.
If $f$
is a function from the set of all families of
configurations $\bss^{(1)},\dots,\bss^{(k)}\in\{-1,+1\}^n$
into $\R$ such that
$|f(\bss^{(1)},\dots,\bss^{(k)})|
\leq {c}\alpha^2_nq_{\max}$,
then
\begin{equation}
\label{sum-factor}
 2^{-nk}
\sum_{\bss^{(1)},\dots,\bss^{(k)}}
e^{f(\bss^{(1)},\dots,\bss^{(k)})}
=1+o(1)
\end{equation}
as $n\to\infty$.
\end{lemma}

 We will now show that  Lemmas~\ref{lem:comb-bds} and
\ref{lem:sum-factor} combined with \eqref{Approx-Fact} and
\eqref{First-moment-prob} immediately imply Theorem~\ref{thm:main1}.
Indeed, let us choose $\lambda_n$ in such a way that
$\alpha_n=o(\lambda_n)$, $\lambda_n=o(\sqrt n)$ and
$e^{-\lambda_n^2/2}$ decays faster than any power of $n$. Invoking
\eqref{Zm-multi} and using Lemma~\ref{lem:comb-bds} (2) to bound the
sum over configurations which do not satisfy \eqref{const-lan} and
the bounds \eqref{Approx-Fact} and \eqref{First-moment-prob}
 to approximate  the
remaining terms on the right hand side of
\eqref{Zm-multi}, we get the estimate
\begin{equation}
\begin{aligned}
\label{Zm-multi-1}
\mathbb{E}
\Bigl[\prod_{\ell=1}^m
&(Z_n(a_n^{\ell},b_n^{\ell}))_{k_{\ell}}\Bigr]
=(1+o(1)) \Bigl(\prod_{i=1}^k\gamma_{\ell(i)}\Bigr)
{2^{-nk}}\sumpp_{\bss^{(1)} ,\dots, \bss^{(k)}} e^{O(\alpha_n^2
q_{\text{max}}) +o(1)}
\end{aligned}
\end{equation}
where the sum $\sumpp $ runs over
families of linearly
independent configurations
$\bss^{(1)}$, $\dots$, $\bss^{(k)}$ that
satisfy \eqref{const-lan}.
Using Lemma~\ref{lem:comb-bds} (1) to extend the
sum to a sum over all families of configurations
$\bss^{(1)}$, $\dots$, $\bss^{(k)}\in\{-1,+1\}^n$,
we then
refer to the statement
of Lemma~\ref{lem:sum-factor} to complete the proof of
Theorem~\ref{thm:main1}.

The above considerations also indicate
why the asymptotic factorization of the factorial
moments fails if $\alpha_n$ grows faster than
$o(n^{1/4})$.  Indeed, expressing the matrix elements
of $C$ as
$C_{ij}=\delta_{ij}+\tilde q_{ij}$
where $\tilde q_{ij}=q(\bss^{(i)},\bss^{(j)})$
if $i\neq j$ and $\tilde q_{ij}=0$ if $i=j$,
$(\bold x,C^{-1}\bold x)=\sum_j x_j^2
-\sum_{i\neq j}x_i\tilde q_{ij}x_j
+ O(q_{\max}^2\|\bold x\|_2^2)$,
where, as usual, $\|\bold x\|_2$ denotes the
$\ell_2$ norm of $\bold x$.  This
 in turn leads to the more precise
estimate
\begin{equation}
\begin{aligned}
\label{Zm-multi-2}
\mathbb{E}
\Bigl[\prod_{\ell=1}^m
&(Z_n(a_n^{\ell},b_n^{\ell}))_{k_{\ell}}\Bigr]
\\&
=
\Bigl(\prod_{i=1}^k\gamma_{\ell(i)}\Bigr)
{2^{-nk}}\sum_{\bss^{(1)} ,\dots, \bss^{(k)}}
e^{\frac{\alpha_n^2}2\sum_{i\neq j}q_{ij}
+O(\alpha_n^2 q^2_{\text{max}})+o(1)}
+o(1)
\end{aligned}
\end{equation}
for the case where $X_1,\dots,X_n$ are standard normals.
While the  term $\frac{\alpha_n^2}2\sum_{i\neq j}q_{ij}$
is negligible if $\alpha_n=o(n^{1/4})$, it
becomes important as soon as $\alpha_n$ grows like
$n^{1/4}$ or faster, leading to the failure of
asymptotic factorization for the factorial moments.
The details are straightforward but a little tedious
and are given in Section~\ref{sec:Gauss}
for
the Gaussian case, and in Section~\ref{sec:Fast-Grow}
for the non-Gaussian case.

\subsection{Integral representation}
\label{sec:Int-Rep}

As discussed in the preceding subsections,
the convergence of the factorial moments of
$Z_n$ reduces to the proof of the approximate
factorization
formula \eqref{Approx-Fact}.
Following a strategy that was
already used in \cite{borgs:chayes:pittel:01}
and \cite{part1},
our proof
for the case where
$X_1,\dots,X_n$ are not normally distributed
uses an integral representation for the probabilities
on the left hand side of \eqref{Approx-Fact}.

To derive this integral representation
we first express the indicator function $I^{(\ell)}(\bss)$
in terms of the function
$\rect(x)$ defined to be $1$ if $|x|\leq 1/2$ and $0$ otherwise.
Using Fourier inversion and the fact that the Fourier
transform of the function $\rect(x)$ is equal to
$\sinc(f) =  \frac{\sin \pi f}{\pi f} $
this leads to the representation
\eq \label{Iell-int3}
I^{(\ell)}(\bss) = q_{n,\ell}
\int_{-\infty}^{\infty}
 \;
\sinc(f q_{n,\ell} )\cos(2\pi f t_n^{\ell}\sqrt{n}) e^{2\pi i f \sum_{s=1}^n
\sigma_s X_s} df,
\en
where $t_n^{\ell} =
({a_n^{\ell}+b_n^{\ell}})/2$ denotes the center of the interval
$[a_n^{\ell},b_n^{\ell}]$, and
$q_{n,\ell}=\gamma_\ell\xi_n\sqrt{n}$.
Taking the expectation of both sides of \eqref{Iell-int3}
and exchanging the expectation
and the integral on the right hand side (the justification of this
exchange is given by Lemma 3.2 in \cite{part1}), we get
the representation
\begin{equation}
\label{ezn}
\begin{aligned}
\PR\Bigl(E(\bss)\in
[a_n^{\ell},b_n^{\ell}]\Bigr)
&= 2\mathbb{E}[I^{(\ell)}(\bss)]
\\
&= 2 q_{n,\ell}\int_{-\infty}^{\infty}
\; \sinc(fq_{n,\ell})\cos(2\pi f t_n^\ell \sqrt{n})\rhat^n (f) df
\end{aligned}
\end{equation}
where
$
\rhat(f)=\mathbb{E}[e^{2\pi i f X}]
$
is the Fourier transform of $\rho$.

When deriving an integral representation for the
terms on the right hand side of \eqref{Zm-multi},
we will have to take the expectation of a product of
integrals of the form \eqref{Iell-int3}.
Neglecting, for the moment, problems associated with the exchange
of integrations and expectations, this is not difficult.  Indeed,
rewriting the product of integrals as
\begin{equation}
\begin{aligned}
\prod_{j=1}^{k} &\int_{-\infty}^\infty
\sinc(f_j q_{n,\ell(j)})
\cos(2\pi f_j t_n^{\ell(j)} \sqrt{n})
e^{2\pi i f_j \sum_{s=1}^n
\sigma_s X_s} df_j
\\
&= \iiint_{-\infty}^{\infty} \;
\prod_{s=1}^{n}e^{2\pi i v_s X_s}
\prod_{j=1}^{k}\sinc(f_j q_{n,\ell(j)})
\cos(2\pi f_j t_n^{\ell(j)} \sqrt{n})
df_j
\end{aligned}
\end{equation}
where
\begin{equation}
v_s=\sum_{j=1}^k \sigma_s^{(j)} f_j,
\end{equation}
and taking expectations of both sides, we easily arrive at
the representation
\begin{equation}
\label{int3-multi}
\begin{aligned}
\PR\Bigl(E(\bss^{(j)})&\in [a_n^{\ell(j)},b_n^{\ell(j)}]
  \text{ for all } j=1,\dots,k\Bigr)=
\prod_{\ell=1}^m (2q_{n,\ell})^{k_\ell}
\\
&
\times\iiint_{-\infty}^{\infty} \;  \prod_{s=1}^n \rhat(v_s)
\prod_{j=1}^{k} \sinc(f_j q_{n,\ell(j)})
\cos(2\pi f_j t_n^{\ell(j)} \sqrt{n})df_j.
\end{aligned}
\end{equation}
But here a little bit more care is needed to justify
the exchange of
expectation and  integrals.  Indeed, this exchange can
only be justified if
$\bss^{(1)},\dots, \bss^{(k)}$
are linearly independent (see Lemma 3.5 in \cite{part1}),
but luckily, this will be all we need.

\subsection{The SK Model}
\label{sec:SK-Strat}

As in the proof for the {\npp}, we define
$Z_n(a,b)$ to be the number
of points in the energy spectrum that fall into
the interval $[a,b]$, where the energy of a configuration
$\bss$ is now given by \eqref{E-SK}.  Given a family of
non-overlapping intervals $[c_1,d_1]$,
$\dots$, $[c_m,d_m]$ with $c_\ell\geq 0$,
and $d_\ell=c_\ell+\gamma_\ell>c_\ell$ for $\ell=1,\dots,m$,
we now consider the intervals $[a_n^\ell,b_n^\ell]$
with
$a_n^\ell=\alpha + c_\ell \tilde\xi_n$ and $b_n^\ell=\alpha +
d_\ell\tilde \xi_n$,
where $\tilde\xi_n$ is defined in  \eqref{tilde-xi-n}.
Theorem~\ref{thm:SK} now follows immediately from the following
two theorems.

\begin{theorem}
\label{thm:main1-SK}
Let
$\alpha_n=o(n)$ be a sequence of positive real numbers, let
$m$ be a positive integer, and for $\ell=1,\dots, m$,
let $a_n^\ell$ and  $b_n^\ell$ be as above.
For an arbitrary $m$-tuple of positive
integers
$(k_1,..,k_m)$ we then
have
\eq \label{mulfact-SK}
\lim_{n\ra\infty}
\mathbb{E}[\prod_{\ell=1}^m
(Z_n(a_n^{\ell},b_n^{\ell}))_{k_{\ell}}] = \prod_{\ell=1}^m
\gamma_\ell^{k_{\ell}}. \en
\end{theorem}

\begin{theorem}
\label{thm:main2-SK}
There exists a constant $\epsilon_0>0$ such the
following statements hold for all $c\geq 0$ and $\gamma>0$,
all sequences
of positive real numbers $\alpha_n$ with $\alpha_n\leq\epsilon_0n$,
and $a_n,b_n$ of the form
$a_n=\alpha_n+c\tilde\xi_n$, $b_n=\alpha_n+(c+\gamma)\tilde\xi_n$.

(i) $\lim_{n\to\infty}\EX[Z_n(a_n,b_n)]=\gamma$.

(ii) $\EX[(Z_n(a_n,b_n))^3]$ is bounded uniformly in $n$.

(iii) If $\limsup\frac{\alpha_n}n>0$, then $\limsup_{n\to\infty}
\EX[(Z_n(a,b))_2]>\gamma^2$.
\end{theorem}

By now, the proof strategy for these theorems is  straightforward:
As before, the first moment is given as an integral over
the energy density. For the SK model, this density
is given by\eqref{g-SK}, leading to
$\EX[Z_n(a_n^\ell,b_n^\ell)]=\gamma_\ell(1+o(1))$
as long as $\tilde\xi_n=o(1)$.
To analyze the higher factorial moments, we again use
a representation of the form \eqref{Zm-multi}.  Neglecting
for the moment the issue of bounding the sum over atypical
configurations, we will again study the factorization
properties of probabilities of the form
\begin{equation}
\label{P-k-SK}
\PR\Bigl(E(\bss^{(j)})\in [a_n^{\ell(j)},b_n^{\ell(j)}]
  \text{ for all } j=1,\dots,k\Bigr).
\end{equation}
For the SK model, this is even easier than for the {\npp}
with Gaussian noise, since $E(\bss)$ is now itself a Gaussian random
variable, rather than the absolute value of a Gaussian. The joint
distribution of $E(\bss^{(1)})$, $\dots$, $E(\bss^{(k)})$ therefore
has density
\begin{equation}
\label{SK-density-k}
\tilde g^{(k)}(\bold x)=
\frac 1{(2\pi)^{k/2}}\frac 1{\sqrt{\det C}}
\exp\Bigl(-\frac 12(\bold x, C^{-1}\bold x)\Bigr),
\end{equation}
where $C$ is the covariance matrix
\begin{equation}
\label{C-SK}
C_{ij}
=\EX[E(\bss^{(i)})E(\bss^{(j)})]=
n(q(\bss^{(i)},\bss^{(j)}))^2.
\end{equation}
Expanding $C^{-1}_{ij}$ as
$C^{-1}_{ij}=\frac 1n(\delta_{ij}+O(q^2_{\max}))$,
we then get the following analogue of the factorization
formula \eqref{Approx-Fact}:
\begin{equation}
\begin{aligned}
\label{Approx-Fact-SK}
\PR\Bigl(E(\bss^{(j)})&\in [a_n^{\ell(j)},b_n^{\ell(j)}]
  \text{ for all } j=1,\dots,k\Bigr)
  \\
&=\prod_{i=1}^k
\PR\Bigl(E(\bss^{(j)})\in [a_n^{\ell(j)},b_n^{\ell(j)}]
 \Bigr)
e^{O(\alpha_n^2 n^{-1} q^2_{\text{max}})+o(1)}.
\end{aligned}
\end{equation}
In the exponent of the above expression, note that the additional
factor of $n^{-1}$ relative to the analogous expression for the
{\npp} is simply a consequence of the different normalizations of
the energies.  The more significant difference  is the factor of
$q^2_{\text{max}}$ rather than $q_{\text{max}}\textbf{}$, a
consequence of the difference between the covariance matrices for
the two problems. For typical configurations with overlap
$q_{\max}=O(n^{-1/2})$, equation \eqref{Approx-Fact-SK} suggests
that there will be asymptotic factorization if and only if
$\alpha_n=o(n)$. That this is
indeed the case is established in Section~\ref{sec:SK-Proof}.

\section{The {\npp} with Gaussian densities}
\label{sec:Gauss}

In this section we analyze the factorization properties
of the probabilities
\begin{equation}
\label{G.0}
\PR\left(E(\bss^{(j)})\in [a_n^{\ell(j)},b_n^{\ell(j)}]
\text{ for }j=1,\dots,k\right)
\end{equation}
for the case where $X_1,\dots,X_n$ are standard normals. Throughout
this section, we will assume that $\alpha_n=o(\sqrt n)$.

In a preliminary step, we show that it is possible to
approximate the joint density $g^{(k)}(\bold y)$ of
$E(\bss^{(1)})$, $\dots$, $E(\bss^{(k)})$ by its value at
$y_i=\alpha_n$.  To this end we combine the representations
\eqref{H-Gauss-density-k} and \eqref{Gauss-density-k} with the fact
that $C^{-1}$ is bounded uniformly in $n$ if $q_{\max}=o(1)$. For
$y_i\in[a_n^{\ell(i)},b_n^{\ell(i)}]$, we then have
\begin{equation}
g^{(k)}(\bold y)=
g^{(k)}(\boldsymbol\alpha_n)(1+O(\alpha_n\xi_n)+O(\xi_n^2))
=g^{(k)}(\boldsymbol\alpha_n)(1+o(1)),
\end{equation}
implying that
\begin{equation}
\label{G.1}
\PR\left(E(\bss^{(j)})\in [a_n^{\ell(j)},b_n^{\ell(j)}]
\text{ for }j=1,\dots,k\right)
=\Bigl(\prod_{\ell=1}^m
(\xi_n\gamma_\ell)^{k_{\ell}}
\Bigr)
g^{(k)}(\boldsymbol\alpha_n)(1+o(1)).
\end{equation}
Having established this approximation, we now
proceed to prove the  factorization
formula \eqref{Approx-Fact}.

\subsection{Proof of the factorization formula \eqref{Approx-Fact}}

Let us express the right hand side
of \eqref{G.1}
as a sum
over vectors $\bold x\in\R^k$ with $|x_i|=\alpha_n$,
see \eqref{H-Gauss-density-k}.
Recalling the representation
\eqref{Gauss-density-k}, we  expand
$(\bold x,C^{-1}\bold x)$ as
$(\bold x,C^{-1}\bold x)=\|\bold x\|_2^2(1+O(q_{\max}))
=k\alpha_n^2+O(q_{\max}\alpha_n^2)$,
where, as before, $\|\bold x\|_2$ denotes the $\ell_2$-norm of
$\bold x$.  For
$q_{\max}=o(1)$, we further
have $\det C=1+o(1)$, implying that
\begin{equation}
\label{Gauss-Density-Fact}
g^{(k)}(\boldsymbol \alpha_n)
=
\Bigl(\frac 2\pi\Bigr)^{k/2}
e^{-\frac k2\alpha_n^2+O(q_{\max}\alpha_n^2)+o(1)}
=( g(\alpha_n))^k
e^{O(q_{\max}\alpha_n^2)+o(1)}.
\end{equation}
Recalling the approximation
$\PR\Bigl(E(\bss)\in [a_n^{\ell},b_n^{\ell}]\Bigr)=
g(\alpha_n)\gamma_\ell\xi_n(1+o(1))$ established in
Section~\ref{sec:First-Moment-Heur},
the bounds \eqref{G.1} and \eqref{Gauss-Density-Fact}
imply the approximate factorization formula
\eqref{Approx-Fact}.  Note that the only conditions
needed in this derivation {were} the
conditions $\alpha_n=  o(\sqrt n)$ and
$q_{\max}=o(1)$.
Taking into account
Lemma~\ref{lem:comb-bds} (iii), we therefore have
established that \eqref{Approx-Fact} holds whenever
$\alpha_n=o(\sqrt n)$ and
$\bss^{(1)},...,\bss^{(k)}$
obey the condition \eqref{const-lan} for some
$\lambda_n$ of  order $o(\sqrt n)$.

As shown in Section~\ref{sec:Typ-Fact-Heur}, the approximate
factorization formula \eqref{Approx-Fact}
immediately leads to Poisson convergence if $\alpha_n=o(n^{1/4})$.
To establish that this convergence fails for faster growing
$\alpha_n$, requires a little bit more work.  This is done
in the next subsection.

\subsection{Proof of Theorem~\ref{thm:main2}}

We start again from \eqref{G.1},  but this time
we expand the inverse of $C$ a little further.
Explicitly, expressing the matrix elements
of $C$ as
$C_{ij}=\delta_{ij}+\tilde q_{ij}$
where $\tilde q_{ij}=q(\bss^{(i)},\bss^{(j)})$
if $i\neq j$ and $\tilde q_{ij}=0$ if $i=j$,
we clearly have
\begin{equation}
\begin{aligned}
(\bold x,C^{-1}\bold x)
&=\sum_j x_j^2
-\sum_{i\neq j}x_i\tilde q_{ij}x_j
+ O(q_{\max}^2\|\bold x\|_2^2)
\\
&=k\alpha_n^2
-\sum_{i\neq j}x_i \tilde q_{ij}x_j
+ O(q_{\max}^2\alpha_n^2)
\end{aligned}
\end{equation}
whenever $q_{\max}=o(1)$ and $|x_i|=\alpha_n$
for all $i$.  With the help of
\eqref{G.1}, \eqref{xi-n}, \eqref{H-Gauss-density-k}
and \eqref{Gauss-density-k},
this implies that
\begin{equation}
\label{G.2}
\begin{aligned}
\PR&\left(E(\bss^{(j)})\in [a_n^{\ell(j)},b_n^{\ell(j)}]
\text{ for }j=1,\dots,k\right)
\\
&\qquad=\Bigl(\prod_{\ell=1}^m
\gamma_\ell^{k_{\ell}}
\Bigr)
2^{-nk}
\sum_{\bes x_1,\dots x_k:\\ x_i=\pm \alpha_n\es}
\exp\Bigl(\frac 12\sum_{i\neq j}x_i \tilde q_{ij}x_j
+ O(q_{\max}^2\alpha_n^2)+o(1)\Bigr).
\end{aligned}
\end{equation}

Let us now
choose  $\lambda_n$ in
such a way that $\alpha_n=o(\lambda_n)$,
$\lambda_n=o(\sqrt n)$ and $e^{-\lambda_n^2/2}$
decays faster than any power of $n$.
Combining \eqref{G.2} with the representation
\eqref{Zm-multi} and Lemma~\ref{lem:comb-bds} (2),
we then have
\begin{equation}
\begin{aligned}
\label{G.3}
\mathbb{E}
\Bigl[\prod_{\ell=1}^m
&(Z_n(a_n^{\ell},b_n^{\ell}))_{k_{\ell}}\Bigr]
=
\Bigl(\prod_{i=1}^k\gamma_{\ell(i)}\Bigr)
{2^{-nk}}\sumpp_{\bss^{(1)} ,\dots, \bss^{(k)}}
2^{-k}
\\
&\times
\sum_{\bes x_1,\dots x_k:\\ x_i=\pm \alpha_n\es}
\exp\Bigl(\frac 12\sum_{i\neq j}x_i \tilde q_{ij}x_j
+ O(q_{\max}^2\alpha_n^2)+o(1)\Bigr)
+o(1)
\end{aligned}
\end{equation}
where the sum $\sumpp $ runs over
families of linearly
independent configurations
$\bss^{(1)}$, $\dots$, $\bss^{(k)}$ that
satisfy \eqref{const-lan}.
Next we claim that we can extend this sum to a sum
over all families of configurations
$\bss^{(1)}$, $\dots$, $\bss^{(k)}$ at the cost
of an additional additive error $o(1)$.
Indeed,
by Lemma~\ref{lem:comb-bds} (1) and (3), the
number of configurations
$\bss^{(1)}$, $\dots$, $\bss^{(k)}$ that are linearly
dependent or violate the bound \eqref{const-lan} is
bounded by a constant times $2^{nk}e^{-\lambda_n^2/2}$.
Since the terms in the above sum are all bounded
by $e^{O(\alpha_n^2)}=e^{o(\lambda_n^2)}$, the extension
from the sum $\sumpp$ to a sum over all families of
configurations $\bss^{(1)}$, $\dots$, $\bss^{(k)}$
indeed only introduces an additive error $o(1)$.
Observing finally that the exponent
$\sum_{i\neq j}x_i \tilde q_{ij}x_j$ is invariant under
the transformation $x_i\to |x_i|=\alpha_n$ and
$\bss^{(i)}\to \text{sign} (x_i)\bss^{(i)}$, we obtain
the approximation \eqref{Zm-multi-2}, which we restate
here as
\begin{equation}
\begin{aligned}
\label{G.4}
\mathbb{E}
\Bigl[\prod_{\ell=1}^m
&(Z_n(a_n^{\ell},b_n^{\ell}))_{k_{\ell}}\Bigr]
\\&
=
\Bigl(\prod_{i=1}^k\gamma_{\ell(i)}\Bigr)
{2^{-nk}}
\sum_{\bss^{(1)} ,\dots, \bss^{(k)}}
\exp\Bigl(\frac {\alpha_n^2}2\sum_{i\neq j} \tilde q_{ij}
+ O(q_{\max}^2\alpha_n^2)+o(1)\Bigr)
+o(1).
\end{aligned}
\end{equation}
With the approximation \eqref{G.4} in hand,
the second statement
of Theorem~\ref{thm:main2} now follows immediately
from the following lemma.

\begin{lemma}
\label{lem:G1}
Let $C<\infty$, let $\beta_n=o(n)$, let
$\EX_k$ denote expectation with respect to the uniform
measure on all families of configurations
$\bss^{(1)} ,\dots, \bss^{(k)}$, and let $R$
denote a function on families of configurations
such that
$|R(\bss^{(1)} ,\dots, \bss^{(k)})|
\leq Cq_{\max}^2$. Then

\begin{equation}
\label{lemG.1} \EX_k\Bigl[\exp\Bigl(\beta_n\Bigl(\sum_{i\neq j}
\tilde q_{ij}+R\Bigr)\Bigr)\Bigr]=
\exp\Bigl(\frac{k(k-1)}n\beta_n^2+ O(\beta_n^3n^{-2}) + o(1) \Bigr).
\end{equation}
\end{lemma}

\begin{proof}
The complete proof of the lemma will be given
in Section~\ref{sec:lemG1-proof}.  Here we only show
that the leading term behaves as claimed, namely
\begin{equation}
\label{lemG.2}
\EX_k\Bigl[\exp\Bigl(\beta_n\sum_{i\neq j} \tilde q_{ij}\Bigr)\Bigr]=
\exp\Bigl(\frac{k(k-1)}n\beta_n^2+ O(\beta_n^3n^{-2})
\Bigr).
\end{equation}
To this end, we observe that
\begin{equation}
\sum_{i\neq j}\tilde q_{ij}=
\sum_{i,j}q(\bss^{(i)},\bss^{(j)})
- k
=\frac 1n
\sum_{s=1}^n\Bigl(\sum_{i=1}^k\sigma_s^{(i)}\Bigr)^2
-k.
\end{equation}
As a consequence, we have
\begin{equation}
\begin{aligned}
\EX_k\Bigl[
e^{\beta_n\sum_{i\neq j}\tilde q_{ij}}
\Bigr]
&=e^{-k\beta_n}
\EX_k\Bigl[
\prod_{s=1}^n
\exp\Bigl(\frac {\beta_n}n
\Bigl(\sum_{i=1}^k\sigma_s^{(i)}\Bigr)^2\Bigr)
\Bigr]
\\
&=
e^{-k\beta_n}
\tilde\EX_k\Bigl[
\exp\Bigl(\frac {\beta_n}n
\Bigl(\sum_{i=1}^k\delta_i\Bigr)^2\Bigr)
\Bigr]^n,
\end{aligned}
\end{equation}
where $\tilde\EX_k$ denotes expectation with respect
to the uniform measure on all configurations
$\bsd=(\delta_1,\dots,\delta_k)\in
\{-1,+1\}^k$.  Next we expand the expectation on
the right hand side into a power series in $\beta_n/n$
(recall that we assumed $\beta_n=o(n)$).
Using that the expectation of $(\sum_i\delta_i)^2$
is equal to $k$, while the expectation of
$(\sum_i\delta_i)^4$ is equal to $3k^2-2k$, this gives
\begin{equation}
\tilde\EX_k
\Bigl[
\exp\Bigl(\frac {\beta_n}n
\Bigl(\sum_{i=1}^k\delta_i\Bigr)^2\Bigr)
\Bigr] =
\exp\Bigl(
\frac {\beta_n}n k
+
\frac {\beta_n^2}{n^2}
k(k-1)
+
O(\beta_n^3n^{-3})
\Bigr)
\end{equation}
and hence
\begin{equation}
\begin{aligned}
\label{F1-bd}
\EX_k\Bigl[
e^{\beta_n\sum_{i\neq j}\tilde q_{ij}}
\Bigr]
&=\exp\Bigl(
\frac {k(k-1)}{n}
\beta_n^2
+
O(\beta_n^3n^{-2})
\Bigr).
\end{aligned}
\end{equation}
\end{proof}

\begin{remark}
\label{rem:overlap}
For the special case where $\alpha_n$ grows like $n^{1/4}$,
Theorem~\ref{thm:main2} implies that the density of the
process $N_n(t)$ converges to one, while the process itself
does not converge to Poisson.   Another interesting consequence
of our proof is the following: Consider the rescaled overlap
$Q_{n,t}$ of two typical configurations contributing to $N_n(t)$
(see Section~\ref{sec:npp-results} for the precise
definition of $Q_{n,t}$).  Then $Q_{n,t}$
converges in distribution to the superposition of two Gaussians with mean
$\pm\kappa^2$, where $\kappa=\lim_{n\to\infty}\alpha_n n^{-1/4}$,
\eq
\lim_{n\to\infty}\PR(Q_{n,t}\geq y)
=\frac 12\frac{1}{\sqrt {2\pi}}
\int_{y}^\infty \frac{\displaystyle
e^{-\frac 12(x- \kappa^2)^2}
+
e^{-\frac 12(x+  \kappa^2 )^2}}2dx.
\en
This follows again from \eqref{G.2}; in fact, now we only need this
formula for $k=2$, where the sum over $\bold x$ just gives a factor
$\cosh(\alpha_n^2 q(\bss,\tilde\bss))$.  This factor is responsible
for the shift of $  \pm\kappa^2 $ in the limiting
distribution of $ Q_{n,t} =n^{1/2}q(\bss,\tilde\bss)$.
\end{remark}

\section{The {\npp} with general distribution}
\label{sec:Poisson-gen}

In this section we prove
the first moment estimate \eqref{First-moment-prob}
and the factorization formula
\eqref{Approx-Fact} for arbitrary distributions $\rho$
obeying the assumptions of Theorem~\ref{thm:growing},
see Propositions~\ref{prop:first-mom} and
\ref{prop:k-factor} below.
 As discussed in
Section~\ref{sec:Proof-Strat}, this immediately gives
Theorem~\ref{thm:main1}. We also show in
Section~\ref{sec:Fast-Grow} how the proof of non-convergence can
be generalized from the Gaussian case to the general distributions
considered in Theorem~\ref{thm:growing}.

\subsection{Properties of the Fourier transform}
\label{sec:FT-Prop}

Throughout this section, we
will use several properties of the Fourier transform in
$\rhat$ which we summarize in this subsection.

{\renewcommand{\labelenumi}{(\roman{enumi})}
\begin{enumerate}

\item\label{rhat-Prop1}
For any $n \geq n_o$,
where $n_0$ is the solution
of $\frac{1}{1+\epsilon}+ \frac{1}{n_o} = 1$
with $\epsilon$ as in \eqref{dist}, we have
\begin{equation}
\label{bdrhat1}
 \int_{-\infty}^{\infty} |\rhat(f)|^{n} \leq \int_{-\infty}^{\infty}
|\rhat(f)|^{n_0} = C_0 < \infty.
\end{equation}

\item\label{rhat-Prop2}
There exists $\mu_0>0$ such that
$\rhat(f)$ is analytic in the region $|\IM f|<\mu_0$.

\item\label{rhat-Prop4}
For any $\mu_1>0$ there exists   $c_1> 0$ and $ \mu_2>0$ and
such that
\begin{equation}
\label{bdrhat2}
|\rhat(f)|\leq e^{-c_1}
\end{equation}
whenever $|\IM f|\leq \mu_2$
and $|\RE f| \geq \mu_1$.
\end{enumerate}
} \noindent These properties easily follow from our
assumptions on the density $\rho$. The bound \eqref{bdrhat1} is a
direct consequence of \eqref{dist}. Analyticity of $\rhat$ in a
neighborhood of the origin implies existence of exponential moments
and this in turn implies analyticity in a strip about the real line.
To see that the bound \eqref{bdrhat2} is true, we first observe that
it obviously holds when $f$ is real.  From the fact that
$d\rhat(z)/dz$ is bounded uniformly in $z$ as long as the imaginary
part of $z$ is small enough, we can choose $\mu_2$ in such a way
that the bound \eqref{bdrhat2} extends to $|\IM f|\leq \mu_2$ (with
$c_1$ replaced by a slightly smaller constant, which we again call
$c_1$).

\subsection{First Moment}
\label{section:firstmoment}

In this section we establish the representation
\eqref{gn.largedev} for the density $g_n(\cdot)$,
see Proposition~\ref{prop:first-mom} below.
As explained in Section~\ref{sec:First-Moment-Heur},
this representation immediately give the first
moment estimate \eqref{First-moment-prob} for
$\alpha_n=o(\sqrt n)$.

We will start from the integral representation
\begin{equation}
\label{gn-int-rep}
g_n(\alpha)= 2 \sqrt n\int_{-\infty}^{\infty}
\cos(2\pi f \alpha\sqrt{n})\rhat^n (f) df,
\end{equation}
This formula  can easily be derived by first
expressing the density of $H(\bss)$ in terms of
the $n$-fold convolution of $\rho$ with itself,
using Fourier inversion to express this density
as an integral over $\rhat^n$, and then summing over
the two possible choices for the sign of $H(\bss)$ given
$E(\bss)$ (alternatively, the formula can be derived
from \eqref{ezn} by sending $\gamma_\ell$ to $0$).

In order to determine the asymptotic behavior of $g_n(\alpha)$, one
might want to expand $\rhat^n (f)$ about its maximum, i.e.~about
$f=0$. While this strategy works well for bounded $\alpha_n$, it
needs to be modified in the case of growing $\alpha_n$. Here we will
use the method of steepest descent, a method which was first
introduced into the analysis of density functions by a paper of
Daniels \cite{daniels:54}, though of course  the ideas  go back to
Laplace.

Before explaining this further, let us first note that,
asymptotically, the integral in \eqref{gn-int-rep} can be restricted
to a bounded interval about zero.  Indeed, let $\mu_1>0$ be an
arbitrary constant, let $\mu_{ 2}$ be as in \eqref{bdrhat2},
and let $n_0$ be as in \eqref{bdrhat1}.  For $n\geq n_0$ and $|f|
\geq \mu_1$, we then have $|\rhat^n (f)| \leq
|\rhat^{n_0}(f)|e^{-c_1(n-n_0)}$. As a consequence, the contribution
to the integral \eqref{gn-int-rep} from the region $\abs{f} \geq
\mu_1$  is bounded by a constant times $\sqrt n e^{-c_1n}$, giving
the estimate

\begin{equation}
\label{fineq2} g_n(\alpha) =2\sqrt n \int_{-\mu_1}^{\mu_1} \;
\cos(2\pi f \alpha \sqrt{n})\rhat^n (f) df +O(\sqrt n e^{-c_1n}).
\end{equation}
Next we observe that both the range of integration and the function
$f\mapsto\rhat(f)$ are invariant under the change $f\to -f$,
implying that we can rewrite the integral as
\begin{equation}
\label{int-before-shift}  2\sqrt n
\int_{-\mu_1}^{\mu_1}e^{2\pi i f \alpha_n \sqrt{n}}\rhat^n (f)df =
2\sqrt n \int_{-\mu_1}^{\mu_1} e^{-n(F(f)-2\pi
i\frac{\alpha_n}{\sqrt n}f)} df
\end{equation}
where $F(f)$ is defined by
\begin{equation}
\rhat(f)=e^{-F(f)}.
\end{equation}

As defined earlier, let $\mu_0$ be such that $\rhat(f)$ is
analytic in the region $|\IM f|<\mu_0$. Further let
$0<\eta\leq\mu_0$, and let $\mathcal C$ be the path in the complex
plane obtained by concatenating the three line segments that join
the point $-\mu_1$ to the point ${-\mu_1+i\eta}$, the point
${-\mu_1+i\eta}$ to the point ${\mu_1+i\eta}$, and the point
${\mu_1+i\eta}$ to the point ${\mu_1}$, respectively.
 Let $\mathcal G$ be the region bounded by
$\mathcal C$ and the  line segment from $-\mu_1$ to $\mu_1$.  Then
the function $f\mapsto \rhat(f)$ and hence the integrand in
\eqref{int-before-shift} is analytic in $\mathcal G$, implying that
the integral on the right hand side of \eqref{int-before-shift} is
equal to the integral over $\mathcal C$. Our next lemma states that
the contribution of the first and third line segment to this
integral is negligible, effectively allowing us to ``shift the
path of integration''  into a parallel line segment in
 the complex
plane.

\begin{lemma}
\label{lem:shift}
Given $\mu_1>0$, there are constants $c_1>0$ and $\mu_2>0$ such that
\begin{equation}
\label{int-after-shift0}
g_n(\alpha_n) =2\sqrt n
\int_{-\mu_1+i\eta}^{\mu_1+i\eta}
e^{-n(F(f)-2\pi i\frac{\alpha_n}{\sqrt n} f)} df +O(\sqrt n e^{-c_1n})
\end{equation}
provided  $\alpha_n \geq 0$ and $0\leq\eta\leq\mu_2$.
\end{lemma}

\begin{proof}
Taking into account
\eqref{fineq2}, we need to analyze the integral
on the right of \eqref{int-before-shift}
which in turn is equal to the integral over the path
$\mathcal C$ defined above,
provided we choose $\mu_2\leq \mu_0$.

Consider the integral
\begin{equation}\int_{-\mu_1}^{-\mu_1+i\eta}
e^{-n(F(f)-2\pi i\frac{\alpha_n}{\sqrt n} f)} df.
\end{equation}
For $f$ in the line segment joining ${-\mu_1}$ to ${-\mu_1+i\eta}$,
the imaginary part of $f$ is non-negative, implying that $|e^{2\pi i
n\frac{\alpha_n}{\sqrt n} f}|\leq 1$. Due to property
\eqref{bdrhat2} of the Fourier transform  $\rhat$, we furthermore
have that $|\rhat(f)|=|e^{-F(f)}|\leq e^{-c_1}$ on this line
segment.  As a consequence, the above integral is bounded by $\eta
e^{-c_1 n}=O(e^{-c_1n}) $.
 In a similar way, the integral over the
third line segment is bounded by
$O(e^{-c_1n})$.
Taking into account the multiplicative factor
of $\sqrt n$ in \eqref{fineq2} and the fact that the
error term in  \eqref{fineq2} is bounded by $O(\sqrt ne^{-c_1n})$,
this proves
\eqref{int-after-shift0}.
\end{proof}

Note that the above lemma holds for arbitrary $\eta\geq 0$.
As usual when applying the method of steepest descent,
the value of $\eta$ will be chosen in such a way that
the asymptotic analysis of the integral on the right
becomes as simple as possible.  Here this amounts to
requiring that at $f=i\eta$, the first derivative of
the integrand is zero.  In other words, we will choose
$\eta$ as the solution of the equation
\begin{equation}
\label{eta-def}
F'(i\eta)=2\pi i\frac{\alpha_n}{\sqrt n},
\end{equation}
where $
F'(z)=\frac{dF(z)}{dz}
$ denotes the complex derivative of $F$.
This leads to the following proposition.

\begin{proposition}
\label{prop:first-mom}
There is an even function $G(x)$ which is real
analytic in a neighborhood of zero such that
\begin{equation}
\label{gn.largedev-rep}
g_n(\alpha_n)=\sqrt{\frac 2\pi}e^{-nG(\alpha_nn^{-1/2})}(1+o(1))
\end{equation}
whenever $\alpha_n=o(\sqrt n)$.  In a neighborhood of zero,
$G(x)$ can be expanded as
\begin{equation}
\label{G.bound-rep}
G(x)=\frac {x^2}2+O(x^4).
\end{equation}
For $\alpha_n=o(\sqrt n)$ and $a_n^\ell$, $b_n^\ell$ and
$\gamma_\ell$ as in Theorem~\ref{thm:main1} we therefore have
\begin{equation}
\label{Prob-asympt}
\PR\Bigl(E(\bss)\in [a_n^{\ell},b_n^{\ell}]\Bigr)=
2^{-(n-1)}\gamma_\ell(1+o(1)).
\end{equation}
\end{proposition}

\begin{proof}

We first argue that the equation \eqref{eta-def} has a unique solution
which can be expressed as an analytic function of $x=\alpha_n/\sqrt n$.
Consider thus the equation
\begin{equation}
\label{eta-x-def}
F'(i\eta)=2\pi i x.
\end{equation}
Since
 $\rhat$ is an even function which is
analytic in a neighborhood of zero
with $\rhat(0)=1$, the function $F$ is even and analytic
in a neighborhood of zero as well.  Taking into account
that the second moment of $\rho$ is one, we now expand
$F(f)$ as
\begin{equation}
\label{F-expan}
F(f)=\frac 12 (2\pi f)^2
+O(f^4)
\end{equation}
and $F'(f)$ as
\begin{equation}
\label{F'-expan}
F'(f)=(2\pi )^2 f
+O(f^3).
\end{equation}
By the implicit function theorem, the equation \eqref{eta-x-def} has
a unique solution $\eta(x)$ in a neighborhood of zero, and $\eta(x)$
is an analytic function of $x$. Taking into  account that $F'$ is
odd, we furthermore have that $\eta(x)$ is an odd function which is
real analytic in a neighborhood of zero, and expanding $\eta(x)$ as
$\eta(x)=\frac 1{2\pi}x+O(x^3)$ we see that $\eta(x)\geq 0$ if
$x\geq 0$ is sufficiently small.

Having established the existence and uniqueness of $\eta(x)$ for
small enough $x$, we are now ready to analyze the integral in
\eqref{int-after-shift0} for $\eta=\eta(x)$ and $x=\alpha_n/\sqrt
n=o(1)$. To this end, we rewrite the integral as
\begin{equation}
\label{int-after-shift}
\begin{aligned}
\int_{-\mu_1}^{\mu_1}
&e^{-n\bigl(F(i  \eta(x) +f)
+2\pi x(\eta(x) -if)\bigr)}
df
= e^{-n  G(x) }
\int_{-\mu_1}^{\mu_1} \;
e^{-n\tilde F(f)} df
\end{aligned}
\end{equation}
where
\begin{equation}
\label{Gn-def}
G(x) =F(i  \eta(x) )+2\pi x  \eta(x) ,
\end{equation}
 and
\begin{equation}
\tilde F(f)=
F(i  \eta(x) +f)-F(i  \eta(x) )-2\pi i x f.
\end{equation}
Next we
would like to show that the integral on the right hand side of
\eqref{int-after-shift} can be restricted to $|f|\leq \log n/\sqrt n$.
To this end, we
expand $\tilde F(f)$  about $f=0$.  Taking into account the
fact that the first derivative at $f=0$ vanishes by the
definition of $ \eta(x) $, we get
\begin{equation}
\begin{aligned}
\tilde F(f)
&=\frac 12F^{(2)}(i\eta(x))f^2
+\frac 16 F^{(3)}(i\eta(x))f^3 +O(f^4).
\end{aligned}
\end{equation}
Using again that $F$ is an even function of its argument, we conclude
that $F^{(2)}(i\eta(x))$ is real, while
$F^{(3)}(i\eta(x))$ is purely imaginary, so that
\begin{equation}
\Re e\,\tilde F(f)
=\frac 12F^{(2)}(i\eta(x))f^2+ O(f^4).
\end{equation}
Since  $  \eta(x)=O(x) =o(1)$
and $F^{(2)}(i  \eta(x) )=4\pi^2+O(  \eta(x) ^2)$,
we have that $\Re e\,\tilde F(f)\geq f^2$
provided $\mu_1$ is sufficiently small and
$n$ is sufficiently large. As a consequence, we
get that
\begin{equation}
\label{int-after-shift1}
\int_{-\mu_1}^{\mu_1} \;
e^{-n\tilde F(f)} df
=\int_{-n^{-1/2}\log n}^{n^{-1/2}\log n } \;
e^{-n\tilde F(f)} df
+O\Bigl(\frac 1{\sqrt n}e^{-\log^2 n}\Bigr).
\end{equation}
For $|f|\leq \log n/\sqrt n$, we now expand
\begin{equation}
\label{exp-after-shift}
\begin{aligned}
e^{-n\tilde F(f)}
&=
\exp\Bigl(
- \frac n2F^{(2)}(i  \eta(x) )f^2
+O(n f^3)\Bigr)
\\
&=\exp\Bigl(
- \frac n2F^{(2)}(i  \eta(x) )f^2
\Bigr)\Bigl(1+O(nf^3)\Bigr),
\end{aligned}
\end{equation}
leading to the approximation
\begin{equation}
\label{int-after-shift2}
\begin{aligned}
\int_{-n^{-1/2}\log n}^{n^{-1/2}\log n } \;
e^{-n\tilde F(f)} df
&=\sqrt{\frac{2\pi }{{nF^{(2)}(i  \eta(x) ) }}}\bigl(1+O(n^{-1/2})\bigr)
=\frac{1}{\sqrt{2\pi n}}
 \bigl(1+o(1)\bigr).
\end{aligned}
\end{equation}
Combined with
\eqref{int-after-shift0},
\eqref{int-after-shift} and \eqref{int-after-shift1}
this proves   \eqref{gn.largedev-rep}.

Next, we show
that $G(x)$ is an even function which is real
analytic in a neighborhood of zero and obeys the bound
\eqref{G.bound-rep}.  But this is almost obvious by now.
Indeed, combining the fact that $F$ is an even function
which is real analytic in a neighborhood of zero with the
fact that $\eta(x)$ is an odd function which is real
analytic in a neighborhood of zero, we see that $G(x)$
is an even function which is real
analytic in a neighborhood of zero.
With the help of \eqref{F-expan}, the bound
\eqref{G.bound-rep} finally follows by inserting the
expansion $\eta(x)=\frac 1{2\pi} x+O(x^3)$ into
the definition \eqref{Gn-def} of $G$.

As we already argued in Section~\ref{sec:First-Moment-Heur},
the bound \eqref{Prob-asympt} finally follows immediately from
the remaining statements of the proposition.
\end{proof}

\subsection{Higher Moments}

In order to establish convergence of the higher factorial moments,
we start from the integral representation \eqref{int3-multi}.
Recalling the definition \eqref{n-delta}, let
\begin{equation}
n_{\min}=n_{\min}(\bss^{(1)},...,\bss^{(u)})
=\min\{n_{\bsd}\colon\bsd\in \{-1,+1\}^u\}.
\end{equation}
In a first step, we  show that under the condition that
$n_{\min}\geq n_0$ where $n_0$ is the solution
of $\frac{1}{1+\epsilon}+ \frac{1}{n_o} = 1$
with $\epsilon$ as in \eqref{dist},
the integral in
\eqref{int3-multi}
is well
approximated by an integral
over a bounded domain, with the product of the $\sinc$ factors replaced by one
and the various midpoints $t_n^\ell$ replaced by $\alpha_n$.

\begin{lemma}
\label{lem:origin-u} Given $\mu_1>0$ there exists a constant $c_1>0$
such that for $\alpha_n=o(\sqrt n)$ and
$n_{\min}=n_{\min}(\bss^{(1)},..,\bss^{(k)})\geq n_0$  we have
\begin{equation}
\label{int3-u}
\begin{aligned}
\PR&\Bigl(E(\bss^{(j)})\in [a_n^{\ell(j)},b_n^{\ell(j)}]
  \text{ for } j=1,\dots,k\Bigr)=
\prod_{\ell=1}^m (2q_{n,\ell})^{k_\ell}
\\
& \times\bigg( \iiint_{-\mu_1}^{\mu_1} \;  \prod_{\bsd}
\rhat({\mathbf f}\cdot\bsd)^{n_\bsd} \prod_{j=1}^{k} \cos(2\pi f_j
\alpha_n \sqrt{n})df_j+O(e^{-c_1n_{\min}})+ O(\sqrt
n\xi_n)\bigg).
\end{aligned}
\end{equation}
 Here the product over $\bsd$ runs over all $\bsd\in\{-1,+1\}^k$
and ${\mathbf f}\cdot\bsd$ stands for the scalar product
$\sum_jf_j\delta_j$.
\end{lemma}

\begin{proof}
Consider
the  integral
on the right hand side of \eqref{int3-multi}.
In a first step, we will consider
the contribution of the region where
$|f_j|>\mu_1$ for at least one $j$ and show that it is negligible.
To this end, it is clearly enough to bound
the contribution of the region where
$\sum_{j=1}^k|f_j|>\mu_1$.  But if
$\sum_{j=1}^k|f_j|>\mu_1$, then there is a
$\bsd\in\{-1,+1\}$ such that
$|\sum_{j=1}^k\delta_jf_j|>\mu_1$.
Since $n_\bsd\geq 1$ for all $\bsd\in\{-1,+1\}^k$,
we conclude that there exists an $s\in\{1,\dots,n\}$ such that
$|v_s|>\mu_1$.

Thus consider the event that one of the $|v_s|$'s, say $|v_{t_1}|$,
is larger than $\mu_1$. Let $\bsd_1 =
\{ \sigma_{t_1}^{(1)},..,\sigma_{t_1}^{(k)} \}$, and let $\bsd_2,...,\bsd_k$
be vectors such that the rank of the matrix $\Delta$
formed by $\bsd_1,..,\bsd_k$ is $k$.
Let $\{v_{t_2},...,v_{t_k}\}$ be defined by
\begin{equation}
v_{t_i} = \sum_{j=1}^k \delta_i^j f_j.
\end{equation}
Since $\Delta$ has rank $k$, we can
change the variables of integration from $f_j$ to $v_{t_j}$. Let the
Jacobian of this transformation be $J_k$, i.e., let $J_k=|\det \Delta|^{-1}$
where $\det\Delta$ is the determinant of $\Delta$.  Since
 $\Delta$  has entries $\pm 1$
and is non-singular,
we conclude that $|\det \Delta|\geq 1$, implying that $|J_k|\leq 1$.
We therefore may  bound the integral over the region where
where $|v_{t_1}|>\mu_1$  by
\begin{equation}
\label{onetwo-1}
\begin{aligned}
 &\left| \iiint_{-\infty}^{\infty} \int_{\abs{v_{t_1}} >
\mu_1} \; J_k \prod_{s=1}^n \rhat(v_s)
\prod_{j=1}^k \sinc(f_j q_{n,\ell(j)})\cos(2\pi f_j t_n^{\ell(j)}  \sqrt{n})  dv_{t_j}
\right|  \\
& \leq \iiint_{-\infty}^{\infty}
\prod_{j=2}^k |\rhat(v_{t_j})|^{n_{\bsd_j}} dv_{t_j}
\times \int_{\abs{v_{t_1}}
>\mu_1} |\rhat(v_{t_1})|^{n_{\bsd_1}} dv_{t_1} \\
& \leq (C_0)^{k-1} \int_{\abs{v_{t_1}} > \mu_1}
|\rhat(v_{t_1})|^{n_{\bsd_1}} dv_{t_1}
\leq C_0^k e^{-c_1(n_{\min}-n_0)}
,
\end{aligned}
\end{equation}
where $C_0$ and $c_1$ are as in \eqref{bdrhat1} and \eqref{bdrhat2}.
Since the number of choices for $\bsd_{t_1}$ is bounded by $2^k$,
the bound \eqref{onetwo-1} implies that the contribution of the
region where at least one of the $|f_j|$'s is larger than $\mu_1$ is
bounded by $O(e^{-c_1n_{\min}})$, with the constant implicit in the
$O$-symbol depending on $k$.

Noting that $q_{n,\ell(j)}=O(\sqrt n\xi_n)$, we finally observe that
in
 the region where $|f_j|\leq\mu_1$ for all $j$, we may expand
$\sinc( q_{n,\ell(j)} f_j)$ as $1+O( \sqrt n\xi_n)$ and $\cos(2\pi
f_j t_n  \sqrt{n}) $ as $\cos(2\pi f_j \alpha_n \sqrt{n}) +O( \sqrt
n\xi_n)$.
   Rewriting the product
$\prod_s\rhat(v_s)$ as
\begin{equation}
\prod_{s=1}^n\rhat(v_s)=
\prod_{\bsd\in\{-1,+1\}^k}
\rhat({\mathbf f}\cdot\bsd)^{n_\bsd},\end{equation}
this gives
\eqref{int3-u}.
\end{proof}

Next we rewrite the integral in \eqref{int3-u} as the average
\begin{equation}
\label{E-H-k}
\begin{aligned}
\iiint_{-\mu_1}^{\mu_1} \;
&\prod_{\bsd}\rhat({\mathbf f}\cdot\bsd)^{n_\bsd}
\prod_{j=1}^k \cos(2\pi f_j \alpha_n  \sqrt{n})  df_j
\\
&=
2^{-k}\sum_{\bold x\in\{-\frac{\alpha_n}{\sqrt n},\frac{\alpha_n}{\sqrt n}\}^k}
\iiint_{-\mu_1}^{\mu_1} \;
e^{2\pi i n {\mathbf f}\cdot{\mathbf x} }
\prod_{\bsd}\rhat({\mathbf f}\cdot\bsd)^{n_\bsd}
\prod_{j=1}^k  df_j
\end{aligned}
\end{equation}
where the sum goes over all sequences
${\mathbf x}=(x_1,\dots,x_k)
\in\{-\frac{\alpha_n}{\sqrt n},\frac{\alpha_n}{\sqrt n}\}^k$
 and ${\mathbf f}\cdot{\mathbf x}$
stands again for the scalar product, ${\mathbf f}\cdot{\mathbf
x}=\sum_jf_jx_j$. In order to analyze the integrals on the right
hand side, we will again use the method of steepest decent. To this
end, we first prove the following analog of Lemma~\ref{lem:shift}.
As before, $\mu_0$ is  a constant such that $\rhat(f)$ is analytic
in the strip $|\IM f|<\mu_0$.

\begin{lemma}
\label{lem:shift-k} Given $\mu_1>0$ there are  constants $c_1>0$ and
$\mu_2\in(0,\mu_0)$ such that the following bound holds whenever
$n_{\min}\geq 2^{-(k+1)}n$ and
 $\eta_1,\dots,\eta_k$
 is sequence of real numbers with
$\sum_j|\eta_j|\leq \mu_2$ and $\eta_j x_j\geq 0$ for all $j$:
\begin{equation}
\label{int-after-shift0-u}
\begin{aligned}
&\iiint_{-\mu_1}^{\mu_1} \;
\prod_{\bsd}\rhat({\mathbf f}\cdot\bsd)^{n_\bsd}
\prod_{j=1}^k e^{2\pi i n x_j f_j } df_j
\\
&\quad= \iiint_{-\mu_1}^{\mu_1} \; e^{2\pi n( i {\mathbf
x}\cdot{\mathbf f} -{\mathbf x}\cdot\boldsymbol\eta)}
\prod_{\bsd}\rhat({\mathbf f}\cdot\bsd+i{\boldsymbol
\eta}\cdot\bsd)^{n_\bsd} \prod_{j=1}^k  df_j +O( e^{-\frac 12
c_1n_{\min}}).
\end{aligned}
\end{equation}
\end{lemma}

\begin{proof}
For $j=1,\dots,k$,
let $\mathcal C_j$ be the path consisting of the three line
segments
which join the point $-\mu_1$ to the point
${-\mu_1+i\eta_j}$,
the point
${-\mu_1+i\eta_j}$ to the point
${\mu_1+i\eta_j}$, and the point
${\mu_1+i\eta_j}$ to
the point ${\mu_1}$,
respectively.
For $j=1,\dots, k$, we replace, one by one, the
integrals over the variables
$f_j$ by integrals over the paths $\mathcal C_j$
and then bound the contribution over the part coming from
the two line segments joining
$-\mu_1$ to
${-\mu_1+i\eta_j}$
and
${\mu_1+i\eta_j}$ to
 ${\mu_1}$,
respectively.  In each of step, we then have to bound integrals of the
form
\begin{equation}
\label{int-shift-est}
\iiint \;
\prod_{\bsd}\rhat({\mathbf f}\cdot\bsd)^{n_\bsd}
\prod_{j'=1}^k e^{2\pi i n x_{j'} f_{j'} } df_{j'}
\end{equation}
where $f_1,\dots,f_{j-1}$ run over the line segments
from ${-\mu_1+i\eta_j}$ to
${\mu_1+i\eta_j}$,
$f_j$ runs either over the the line segment from
$-\mu_1$ to
${-\mu_1+i\eta_j}$
or the line segment from
${\mu_1+i\eta_j}$ to
${\mu_1}$,
and $f_{j+1},\dots,f_k$ run over the interval
$[-\mu_1,\mu_1]$.
To bound these integrals, we  note that
in this domain of integration the real part of $f_j$
has absolute value
$\mu_1$, implying in particular that
\begin{equation}
\mu_1\leq \sum_i |\RE f_i|\leq k\mu_1.
\end{equation}
As a consequence, $|\RE \mathbf f\cdot\bsd|\leq k\mu_1$ for all
$\bsd$, and further there exists at least one $\bsd$ for which
$|\RE\mathbf f\cdot\bsd|\geq \mu_1$. (In fact it is easy to see that
there are at least $2^{k-1}$ such $\bsd$'s.) On the other
hand, the assumption $\sum_j|\eta_j|\leq \mu_2$ implies that $|\IM
\mathbf f\cdot\bsd|\leq \mu_2$  for all $\bsd$.

Consider first a vector $\bsd$ for which
$\mu_1\leq|\RE \mathbf f\cdot\bsd|\leq k\mu_1$,
and assume that $\mu_2\leq\mu_0$ is chosen in such a way
that \eqref{bdrhat2} holds.
Under this assumption, we have
$|\rhat(\mathbf f\cdot\bsd)|\leq e^{-c_1}$, implying that
 the term $\rhat({\mathbf f}\cdot\bsd)^{n_\bsd}$  contributes
 a  factor that is at most $e^{- n_{\min}c_1}$.

 To bound $|\rhat(z)|$ when we only know that
 $|\RE z|\leq k\mu_1$, we use  continuity
 in conjunction
 with the fact that $|\rhat(z)|\leq 1$
 for all real $z$.  Decreasing $\mu_2$, if necessary, we
 conclude that
 $|\rhat(z)|\leq e^{c_12^{-k- 2}}$ whenever  $|\RE z|\leq k\mu_1$
 and  $|\IM z|\leq \mu_2$.  As a consequence, the first
 product in \eqref{int-shift-est} can be bounded by
\begin{equation}
\prod_{\bsd}\Bigl|\rhat({\mathbf f}\cdot\bsd)^{n_\bsd}\Bigr| \leq
e^{- n_{\min}c_1}e^{c_12^{-k- 2}n} \leq e^{- \frac 12 n_{\min}c_1}.
\end{equation}
Since the second product is bounded by one as before, and the
integral only contributes a constant, this proves the lemma.
\end{proof}

Next we have to determine the values of the shifts
$\eta_1,\dots,\eta_k$.  According to the method of steepest decent,
we again choose a saddle point of the integrand. Since the integrand
is now a function of $k$ variables, this now gives a system of $k$
equations for $\boldsymbol\eta=(\eta_1,\dots,\eta_k)$:
\begin{equation}
\label{etak-def}
\sum_{\bsd} \frac {n_\bsd}n\delta_jF'(i\bsd\cdot\boldsymbol\eta)
=2\pi ix_j.
\end{equation}
Assume for a moment that this equation has a unique
solution $\boldsymbol\eta=\boldsymbol\eta(\mathbf x)$.
We then define a function
\begin{equation}
\label{Gk-def}
G_{n,k}(\mathbf x)=
\sum_{\bsd}\frac {n_\bsd}n F(i\bsd\cdot\boldsymbol \eta(\mathbf x))+
2\pi\boldsymbol\eta(\mathbf x)\cdot\mathbf x.
\end{equation}

\begin{lemma}
\label{lem:k-after-shift} Let $0\leq \alpha_n=o(\sqrt n)$, let
$q_{\max}=o(1)$, and let $\bold x\in \R^k$ be such that
$|x_i|=\alpha_n/\sqrt n$. Then the equation \eqref{etak-def} has a
unique solution $\boldsymbol\eta=\boldsymbol\eta(\mathbf x)$,
\begin{equation}
\label{eta-k-asympt}
\eta_i(\mathbf x)=
\Bigl(\frac 1{2\pi} \sum_{j=1}^k C^{-1}_{ij} x_j\Bigr)
\Bigl(1+O\bigl(\frac{\alpha_n^2}n\bigr)\Bigr)
\end{equation}
and $\eta_j(\boldsymbol x) x_j\geq 0$ for all $j\in\{1,\dots,k\}$.
In addition, for sufficiently small $\mu_1$,
\begin{equation}
\label{k-int-after-shift4}
\begin{aligned}
\iiint_{-\mu_1}^{\mu_1} \;&
e^{2\pi n( i {\mathbf x}\cdot{\mathbf f} -{\mathbf x}\cdot\boldsymbol\eta)}
\prod_{\bsd}\rhat({\mathbf f}\cdot\bsd+i{\boldsymbol \eta}\cdot\bsd)^{n_\bsd}
\prod_{j=1}^k  df_j
\\
&=e^{-nG_{n,k}(\mathbf x)}
\biggl(\frac1{2\pi n}\biggr)^{k/2}
\Bigl(1+O(n^{-1/2})+O(\alpha_n^2/n)+O(q_{\max})\Bigr).
\end{aligned}
\end{equation}
\end{lemma}

\begin{proof}

For $\bold x=0$, the equation \eqref{etak-def} is obviously solved
by $\boldsymbol \eta=0$.  To obtain existence and uniqueness of
a solution in the neighborhood of zero, we consider the derivative
matrix of the function on the left hand side,
\begin{equation}
A_{ij}(\boldsymbol\eta)=
i\sum_{\bsd}\frac {n_\bsd}n\delta_i\delta_j F''(i\bsd\cdot\boldsymbol \eta).
\end{equation}
Using the fact that
$F''(f)=(2\pi)^2+O(f^2)$ we may expand ${A}_{ij}(\boldsymbol\eta)$
as
\begin{equation}
\label{A-approx}
{A}_{ij}(\boldsymbol\eta)
=A_{ij}(0)+O(\|\boldsymbol\eta\|_2^2),
\end{equation}
and for $\boldsymbol\eta=0$ we have
\begin{equation}
A_{ij}(0)=(2\pi)^2i\sum_{\bsd} \frac {n_\bsd}n\delta_i\delta_j
=(2\pi)^2 iC_{ij},
\end{equation}
where $C_{ij}$ is the overlap matrix $C_{ij}=q(\bss^{(i)},\bss^{(j)})$.
If the maximal off-diagonal overlap $q_{\max}$ is $o(1)$, the matrix
$C$ is invertible, implying in particular that
$A_{ij}(\boldsymbol\eta)$ is non-singular in a neighborhood of zero.
By the implicit function theorem, we conclude that for $\bold x$
sufficiently small, the equation \eqref{etak-def}  had a unique
solution $\boldsymbol\eta(\bold x)$, and by \eqref{A-approx},
we may expand $\boldsymbol\eta(\bold x)$ as
\begin{equation}
\begin{aligned}
\eta_i(\bold x)
&={2\pi i} \sum_{j=1}^k A^{-1}_{ij}(0) x_j+O(\|\bold x\|_2^3)
\\
&=\frac 1{2\pi} \sum_{j=1}^k C^{-1}_{ij} x_j+O(\|\bold x\|_2^3).
\end{aligned}
\end{equation}
For $q_{\max}=o(1)$, the matrix $C^{-1}$ can be
approximated as $C^{-1}_{ij}=\delta_{ij}+o(q_{\max})$, implying in particular
that for $|x_j|=\alpha_n/\sqrt n$,
the leading term on the right hand side is of order $\alpha_n/\sqrt n$.
As a consequence,
we can convert the additive error into a multiplicative error
$1+O(\alpha_n^2/n)$, giving the desired bound \eqref{eta-k-asympt}.
Using once more that $C^{-1}_{ij}=\delta_{ij}+o(q_{\max})$, the
fact that $x_j\eta_j(\boldsymbol x)\geq 0$ is an immediate corollary
of \eqref{eta-k-asympt}.

We are left with the proof of \eqref{k-int-after-shift4}.
To this end, we rewrite the left hand side as
\begin{equation}
\label{k-int-after-shift5}
\begin{aligned}
\iiint_{-\mu_1}^{\mu_1} \;&
e^{2\pi n( i {\mathbf x}\cdot{\mathbf f} -{\mathbf x}\cdot\boldsymbol\eta)}
e^{-\sum_{\bsd}n_{\bsd}F({\mathbf f}\cdot\bsd+i{\boldsymbol \eta}\cdot\bsd)}
\prod_{j=1}^k  df_j
\\
&\qquad=e^{-nG_k(\mathbf x)}
\iiint_{-\mu_1}^{\mu_1}
e^{-n\tilde G_{n,k}(\bold f)}
\prod_{j=1}^k  df_j
\end{aligned}
\end{equation}
with
\begin{equation}
\tilde G_{n,k}(\bold f)
=\sum_{\bsd}
\frac {n_\bsd}n\bigl(F({\mathbf f}\cdot\bsd+i{\boldsymbol \eta}\cdot\bsd)
-F(i{\boldsymbol \eta}\cdot\bsd)\bigr)
-2\pi  i {\mathbf x}\cdot{\mathbf f}.
\end{equation}
Observing that the derivatives of $\tilde G_{n,k}(\bold f)$ at $\bold f=0$
vanish by the definition of $\boldsymbol\eta(\bold x)$,
we now expand $\tilde G_{n,k}(\bold f)$ as
\begin{equation}
\tilde G_{n,k}(\bold f)=
\sum_{\bsd}
\frac {n_\bsd}n
\Bigl(
\frac 12(\bold f\cdot\bsd)^2
F''(i{\boldsymbol \eta}\cdot\bsd)
+
\frac 1{3!}
(\bold f\cdot\bsd)^3
F^{(3)}(i{\boldsymbol \eta}\cdot\bsd)
+O((\bold f\cdot\bsd)^4)\Bigr).
\end{equation}
Arguing as in the proof of
Proposition~\ref{prop:first-mom} we now have that
\begin{equation}
\RE\tilde G_{n,k}(\bold f)
\geq
\sum_{\bsd}
\frac {n_\bsd}n
(\bold f\cdot\bsd)^2
=\sum_{i,j}f_i C_{ij}f_j
\end{equation}
provided  $\mu_1$ is sufficiently small and $n$ is sufficiently
large. Since the maximal off-diagonal overlap $q_{\max}$  is of
order $o(1)$, we conclude that $\RE\tilde G_{n,k}(\bold f)\geq \frac
12 \|\bold f\|_2^2$ provided $\mu_1$ is sufficiently small and $n$
is sufficiently large. As a consequence, the integral in
\eqref{k-int-after-shift5} is dominated by configurations for which
$\|\bold f\|_2$ is smaller than, say, $\log n/\sqrt n$.  More
quantitatively, we have
\begin{equation}
\label{k-int-after-shift6}
\begin{aligned}
&\iiint_{-\mu_1}^{\mu_1}
e^{-n\tilde G_{n,k}(\bold f)}
\prod_{j=1}^k  df_j
\\
&\quad=
\iiint_{\|\bold f\|_2\leq \log n/\sqrt n}
e^{-n\tilde G_{n,k}(\bold f)}
\prod_{j=1}^k  df_j
+O\Bigl(\iiint_{\|\bold f\|_2\geq \log n/\sqrt n}
e^{-n\|\bold f\|^2/2}
\prod_{j=1}^k  df_j\Bigr)
\\
&\quad= \iiint_{\|\bold f\|_2\leq \log n/\sqrt n} e^{-n\tilde
G_{n,k}(\bold f)} \prod_{j=1}^k  df_j + O\Bigl( e^{-\frac
12\log^2n}\Bigr).
\end{aligned}
\end{equation}
For $\|\bold f\|_2\leq \log n/\sqrt n$ we expand $e^{-n\tilde G_{n,k}(\bold x)}$
as
\begin{equation}
\begin{aligned}
\tilde G_{n,k}(\bold f)
&=
\exp\Bigl(-\frac 12\sum_{\bsd}
\frac {n_\bsd}n
(\bold f\cdot\bsd)^2
F''(i{\boldsymbol \eta}\cdot\bsd)
+O(n\|\bold f\|^3)\Bigr)
\\
&=
e^{-\frac 12\sum_{ij}f_iM_{ij}f_j}
\Bigl(1+O(n\|\bold f\|^3)\Bigr)
\end{aligned}
\end{equation}
where $M$ is the matrix with matrix elements
\begin{equation}
\begin{aligned}
M_{ij}&=\sum_{\bsd}
\frac {n_\bsd}n
\delta_i\delta_j
F''(i{\boldsymbol \eta}\cdot\bsd)
=(2\pi)^2C_{ij}+O(\|\boldsymbol\eta\|^2)
\\
&=(2\pi)^2\delta_{ij}+O(\alpha_n^2/n)+O(q_{\max}).
\end{aligned}
\end{equation}
As a consequence,
\begin{equation}
\iiint_{\|\bold f\|_2\leq \log n/\sqrt n}
e^{-n\tilde G_{n,k}(\bold f)}
\prod_{j=1}^k  df_j
=\Bigl(\frac {1} { 2\pi n}\Bigr)^{\frac k2}
\Bigl(1+O(n^{-1/2})+O(\alpha_n^2/n)+O(q_{\max})\Bigr).
\end{equation}
Combined with \eqref{k-int-after-shift5} and \eqref{k-int-after-shift6},
this implies the desired
bound \eqref{k-int-after-shift4}.
\end{proof}

Lemma~\ref{lem:origin-u}, Lemma~\ref{lem:shift-k},
the relation \eqref{E-H-k},
Lemma ~\ref{lem:shift-k} and Lemma~\ref{lem:k-after-shift}
we now easily prove the following proposition.

\begin{proposition}
\label{prop:k-factor} Let $0\leq \alpha_n=o(\sqrt n)$ and let $0\leq
\lambda_n=o(\sqrt n)$. If $n$ is sufficiently large and
$\bss^{(1)}\dots,\bss^{(k)}$ obey the condition \eqref{const-lan},
then
\begin{equation}
\label{k-int-after-shift9}
\begin{aligned}
\PR&\Bigl(E(\bss^{(j)})\in [a_n^{\ell(j)},b_n^{\ell(j)}]
  \text{ for } j=1,\dots,k\Bigr)
=
\Bigl(\prod_{\ell=1}^m \gamma_\ell^{k_\ell}\Bigr)
\\
&
\times
\biggl(\frac{\xi_n}{\sqrt {2\pi} }\biggr)^{k}
\sum_{\bold x\in\{-\frac{\alpha_n}{\sqrt n},
\frac{\alpha_n}{\sqrt n}\}^k}
e^{-nG_{n,k}(\mathbf x)}
\Bigl(1+o(1)\Bigr).
\end{aligned}
\end{equation}

If we strengthen the  condition $\alpha_n=o(\sqrt n)$
to $\alpha_n=o(n^{1/4})$,
we  have\begin{equation}
\begin{aligned}
\label{Approx-Fact-rep}
\PR\Bigl(E(\bss^{(j)})&\in [a_n^{\ell(j)},b_n^{\ell(j)}]
  \text{ for all } j=1,\dots,k\Bigr)
  \\
&=\prod_{i=1}^k
\PR\Bigl(E(\bss^{(j)})\in [a_n^{\ell(j)},b_n^{\ell(j)}]
 \Bigr)
e^{O(\alpha_n^2 q_{\text{max}})+o(1)}.
\end{aligned}
\end{equation}
\end{proposition}

\begin{proof}
Observe that under the condition \eqref{const-lan},
we have that $n_{\min}=n2^{-k} +o(n)$ and $q_{\max}=o(1)$.
We may therefore use
Lemma~\ref{lem:origin-u}, Lemma~\ref{lem:shift-k},
the relation \eqref{E-H-k},
Lemma ~\ref{lem:shift-k} and Lemma~\ref{lem:k-after-shift}
to conclude that under the conditions of
the proposition, there is a constant $c>0$
such that
\begin{equation}
\label{k-int-after-shift8}
\begin{aligned}
\PR&\Bigl(E(\bss^{(j)})\in [a_n^{\ell(j)},b_n^{\ell(j)}]
  \text{ for } j=1,\dots,k\Bigr)=
\prod_{\ell=1}^m (2q_{n,\ell})^{k_\ell}
\\
& \times 2^{-k}\sum_{\bold x\in\{-\frac{\alpha_n}{\sqrt n},
\frac{\alpha_n}{\sqrt n}\}^k}\bigg( e^{-nG_{n,k}(\mathbf x)}
\biggl(\frac1{2\pi n}\biggr)^{k/2} \Bigl(1+o(1)\Bigr)+O( e^{-cn})
+O( \sqrt n\xi_n)\bigg).
\end{aligned}
\end{equation}
Next we expand $G_{n,k}(\mathbf x)$
with the help of the bounds \eqref{F-expan} and
\eqref{eta-k-asympt}, yielding the approximation
\begin{equation}
\label{G-k-approx}
\begin{aligned}
G_{n,k}(\mathbf x)
&=
-\frac 12\sum_{\bsd}\frac {n_\bsd}n
(2\pi\bsd\cdot\boldsymbol \eta)^2
+
2\pi\boldsymbol\eta\cdot\mathbf x
+O(\|\mathbf x\|_2^4)
\\
&=
-\frac {(2\pi)^2}2(\boldsymbol\eta, C\boldsymbol\eta)
+
2\pi\boldsymbol\eta\cdot\mathbf x
+O(\|\mathbf x\|_2^4)
\\
&=\frac {1}2(\bold x, C^{-1}\bold x)
+O\Bigl(\frac {\alpha_n^4}{n^2}\Bigr).
\end{aligned}
\end{equation}
Note that this implies in particular that $nG_{n,k}(\bold
x)=O(\alpha_n^2)+O({\alpha_n^4}/n)= o(n)$. The leading term on the
right hand side of \eqref{k-int-after-shift8} is therefore much
larger than the additive error terms, which both decay exponentially
in $n$.  These additive error terms can therefore be converted into
a multiplicative error term $(1+o(1))$.  Inserting the value of
$q_{n,\ell} (=\gamma_\ell\xi_n\sqrt{n}) $ and using the fact that
$\sum_\ell k_\ell =k$, this yields the approximation
\eqref{k-int-after-shift9}.

To infer the bound \eqref{Approx-Fact-rep}, we use the assumption
$\alpha_n=o(n^{1/4})$ to expand the factor $(\xi_n/\sqrt{2\pi})^k$
as $2^{-nk}e^{k\alpha_n^2/2}(1+o(1))$, and the factor
$e^{-nG_{n,k}(\mathbf x)}$ as
\begin{equation}
\begin{aligned}
e^{-\frac n2 (\bold x, C^{-1}\bold x)+O(\alpha_n^4/n)} &=e^{- \frac
n2 \|\bold x\|_2^2+O( n\|x\|_2^2 q_{\max})}
\bigl(1+O(\alpha_n^4/n)\Bigr)
\\
&=e^{-\frac k2\alpha_n^2+O(\alpha_n^2 q_{\max})}
\bigl(1+o(1)\Bigr).
\end{aligned}
\end{equation}
Inserting these estimates into \eqref{k-int-after-shift9}
and
taking into account the bound \eqref{Prob-asympt},
this gives \eqref{Approx-Fact-rep}.
\end{proof}

\subsection{Failure of Poisson convergence for
faster growing $\boldsymbol\alpha_{\bold n}$}
\label{sec:Fast-Grow}

In this section, we prove the  last statement of
Theorem~\ref{thm:growing} for general distributions.  Throughout
this section, we will assume that $\alpha_n=O(n^{1/{4}})$ and
$\lambda_n=o(\sqrt n)$.  Near the end of the section, we will
specialize to $\alpha_n=\Theta(n^{1/{4}})$, for which we will
prove absence of Poisson convergence.

We start from \eqref{k-int-after-shift9}, but instead of
\eqref{G-k-approx} we use a more accurate approximation for
$G_{n,k}(\mathbf x)$. To this end, we use the fact that
$\rhat(f)=\rhat(-f)$ is analytic in a neighborhood of zero to infer
the existence of a constant $c_4$ such that
\[
F(z)=
\frac {(2\pi )^2}2z^2+
{c_4(2\pi )^4}z^4
+O(|z|^6)
\]
and
\[
F'(z)=
(2\pi )^2z
+4c_4(2\pi )^4z^3
+O(|z|^5).
\]

\begin{remark}
\label{c4} Using the fact that $\mathbb{E}(X^4) \geq
\mathbb{E}(X^2)^2 = 1$ and expanding the $\log \hat{\rho(f)}$ one
can see that $c_4 \leq \frac{1}{12}$, with equality holding if and
only if $X^2 = 1$ with probability one. Thus for
all random variables whose density satisfies assumption \eqref{dist}
we have $c_4 < 1/12$.
\end{remark}

Using  these expressions and the fact that
$C^{-1}_{ij}=\delta_{ij}+O(q_{\max})
=\delta_{ij}+O(\lambda_nn^{-1/2})$  when
$\bss^{(1)}\dots,\bss^{(k)}$ obey the condition \eqref{const-lan},
we then expand the solution of \eqref{etak-def} as
\begin{equation}
\begin{aligned}
\eta_i &=\frac 1{2\pi} (C^{-1}\bold x)_i+\frac {4c_4}{2\pi}
\sum_{\bsd} \frac {n_\bsd}n(C^{-1}\bsd)_i (\bsd,C^{-1}\bold x)^3+
O(\|\bold x\|^5)
\\
&=\frac 1{2\pi} (C^{-1}\bold x)_i+\frac{4c_4}{2\pi} \sum_{\bsd}
\frac {n_\bsd}n\delta_i (\bsd\cdot\bold x)^3+ O(\|\bold
x\|^5)+O(\|\bold x\|^3\lambda_nn^{-1/2}).
\end{aligned}
\end{equation}
In order to analyze the sum over $\bsd$, we use
the condition \eqref{const-lan} to estimate
\[
\sum_{\bsd}\frac{n_\bsd}n\delta_i\delta_j\delta_k\delta_l
=2^{-k}\sum_{\bsd}\delta_i\delta_j\delta_k\delta_l
+O(\lambda_nn^{-1/2}).
\]
The first term is zero unless $i,j,k,l$ are such that either all of
them are equal or are pairwise equal for some pairing of
$i,j,k,l$. Observing that there are $3$ possible ways to pair four
numbers into two pairs of two, this  leads to the estimate
\begin{equation}
\begin{aligned}
\eta_i &=\frac1{2\pi}(C^{-1}\bold x)_i +\frac
{4c_4}{2\pi}x_i\bigl(x_i^2 +3\sum_{k\neq i}x_k^2\bigr)\Bigr)
+O(\|\bold x\|^3\lambda_nn^{-1/2}) +O(\|\bold x\|^5)
\\
&=\frac1{2\pi}(C^{-1}\bold x)_i +
(1+3(k-1))\frac{4c_4\alpha_n^2}{2\pi n}x_i +O(\|\bold
x\|^3\lambda_nn^{-1/2}) +O(\|\bold x\|^5)
\end{aligned}
\end{equation}
where we used the fact that $|x_i|=|x_j|=\alpha_n/\sqrt n$
in the last step.
Inserting this expression into the definition
\eqref{Gk-def} and expanding the result in a similar
way as we expanded $\eta$ above, this leads to the
approximation
\begin{equation}
\begin{aligned}
G_{n,k}&(\mathbf x)= -\frac{(2\pi)^2}2\sum_{\bsd}\frac {n_\bsd}n
(\bsd\cdot\boldsymbol \eta)^2 +c_4 2^{-k}\sum_{\bsd}
(\bsd\cdot\bold x)^4 + 2\pi\boldsymbol\eta\cdot\mathbf x
\\
&\qquad
+O(\|x\|^6)+O(\|x\|^4\lambda_nn^{-1/2})
\\
&=
-\frac{(2\pi)^2}2
(\boldsymbol\eta,C\boldsymbol \eta)
+c_4
k(1+3(k-1))\frac{\alpha_n^4}{n^2}
+
2\pi\boldsymbol\eta\cdot\mathbf x
\\
&\qquad
+O(\|x\|^6)+O(\|x\|^4\lambda_nn^{-1/2})
\\
&= -\frac 12(\bold x,C^{-1}\bold x) -
k(1+3(k-1))\frac{4c_4\alpha_n^4}{ n^2} +{c_4}
k(1+3(k-1))\frac{\alpha_n^4}{n^2}
\\
&\qquad
+(\bold x,C^{-1}\bold x)
+k(1+3(k-1))\frac{4c_4\alpha_n^4}{n^2}+O(\|x\|^6)
\\
&\qquad +O(\|x\|^4\lambda_nn^{-1/2})
\\
&= \frac 12(\bold x,C^{-1}\bold x) +c_4
k(1+3(k-1))\frac{\alpha_n^4}{n^2}
+O\Bigl(\frac{\alpha_n^6}{n^3}\Bigr) +O( \frac{\alpha_n^
4}{n^2}\lambda_nn^{-1/2}).
\end{aligned}
\end{equation}
Using that $\alpha_n=O(n^{1/{4}})$ and   $\lambda_n=o(\sqrt n)$,
this gives
\[
nG_{n,k}(\bold x) =\frac n2(\bold x,C^{-1}\bold x) +
c_4 k(1+3(k-1))\frac{\alpha_n^4}{n} + o(1).
\]
As a consequence, we have that
\begin{equation}
\label{k-int-after-shift7-sharper}
\begin{aligned}
\PR&\Bigl(E(\bss^{(j)})\in [a_n^{\ell(j)},b_n^{\ell(j)}]
  \text{ for } j=1,\dots,k\Bigr)
=
\Bigl(\prod_{\ell=1}^m \gamma_\ell^{k_\ell}\Bigr)
\\
&
\times
\biggl(\frac{\xi_n}{\sqrt {2\pi} }\biggr)^{k}
\sum_{\bold x\in\{-\frac{\alpha_n}{\sqrt n},
\frac{\alpha_n}{\sqrt n}\}^k}
 e^{-\frac n2
(\bold x, C^{-1}\bold x)} e^{- c_4k(1+3(k-1))\frac{\alpha_n^4}{n}}
\Bigl(1+o(1)\Bigr)
\end{aligned}
\end{equation}
whenever $\bss^{(1)}\dots,\bss^{(k)}$ obey the condition
\eqref{const-lan}.

Next we expand $\xi_n$, using that
$\xi_n=(2^{(n-1)}g_n(\alpha_n))^{-1}$ with $g_n(\alpha_n)$ given by
\eqref{gn.largedev-rep}.  To this end, we approximate $\eta(x)$ as
$\eta(x)=\frac x{2\pi}(1+4c_4x^2)+O(x^5)$ and $G(x)$ as
\[
G(x)
=
-\frac 12 (2\pi\eta(x) )^2
+{c_4(2\pi\eta(x) )^4}
+2\pi x\eta(x)+O(x^6)
=
\frac {x^2}2 + {c_4x^4} + O(x^6),
\]
giving the expansion
\begin{equation}
\xi_n=\sqrt{\frac \pi 2}2^{-(n-1)}
\exp\Bigl(\frac 12\alpha_n^2+c_4\frac{\alpha_n^4}n\Bigr)
\bigl(1+o(1)\bigr).
\end{equation}
Inserting this expansion into \eqref{k-int-after-shift7-sharper} we
then continue as in the proof of Theorem~\ref{thm:main2} to get that
\eq \label{mulfact-non-gauss} \mathbb{E}[\prod_{\ell=1}^m
(Z_n(a_n^{\ell},b_n^{\ell}))_{k_{\ell}}] = \Bigl(\prod_{\ell=1}^m
\gamma_\ell^{k_{\ell}}\Bigr)
\exp\Bigl(\frac{\alpha_n^4}{n}\Bigl(\frac{k(k-1)}4 -3c_4
k(k-1)\Bigr)\Bigr) \Bigl(1+o(1)\Bigr). \en
Specializing now to $\alpha_n=\kappa
n^{1/4}$, this implies in particular that all moments of $Z_n(t)$
are bounded, and
\begin{equation}
\label{final-kill} \lim_{n\to\infty}\EX[(Z_n(t))_k]
=t^k\exp\Bigl(\kappa^4(k-1)k\Bigl(1/{4}-3c_4 \Bigr)\Bigr).
\end{equation}
In view of Lemma~\ref{lem:non-poisson} and the fact that $c_4 <
1/12$ by Remark~\ref{c4}, this is incompatible with weak convergence
to a Poisson random variable. This establishes the failure of the
modified REM conjecture for
$\lim_{n\to\infty}\alpha_n^{-1/4}=\kappa>0$.

Note that the original REM conjecture also fails if
$\lim_{n\to\infty}\alpha_n n^{-1/4}=\kappa>0$. Indeed, in this
range, the original scaling \eqref{xi-n} and the modified scaling
differ by the asymptotically constant factor
$e^{-c_4\kappa^4}(1+o(1))$. This makes things worse, since now even
the first moment does not converge to the desired value unless
$c_4=0$,  in which case the original and the modified REM conjecture
remain equivalent for $\alpha_n=O(n^{1/4})$. This completes the
proof of the last statement of Theorem~\ref{thm:growing}.

\begin{remark}
\label{rem:overlap-non-gauss}
It is not hard to see that for
$\alpha_n=\kappa n^{1/4}$ the distribution of the overlap $Q_{n,t}$
converges again to a superposition of two shifted Gaussians, see
Remark~\ref{rem:overlap}, where this was discussed for the case
where $X_1,\dots,X_n$ where standard normals. Indeed, the only
difference with the situation discussed in that remark is the extra
factor of $e^{-c_4(k-1)(1+3k)\kappa^4}$. But this
factor does not depend on the overlap, and hence does not influence
the overlap distribution.
\end{remark}

\section{Analysis of the SK model}
\label{sec:SK-Proof}

For $\alpha_n=O(n^\eta)$ and $\eta<1$, the local REM conjecture
for the SK model has been proved in \cite{bovier:kurkova:05}.
To extend the proof up to the threshold $\alpha_n=o(1)$,
a little bit more care is needed, but given
the analysis of the {\npp} with Gaussian noise
from the Section~\ref{sec:Gauss}, this is still  relatively straightforward,
even though the details vary at several places.
But for $\alpha_n$ of order $n$ we have to be quite careful, since
now several error terms which went to zero before are not
vanishing anymore.  It turns out, however, that for $\alpha_n/n\leq \epsilon_0$
and $\epsilon_0$
sufficiently small, we can at least control the first three moments,
which is enough to prove absence of Poisson convergence
if $\limsup\alpha_n/n>0$.

 We start with the analysis of
the first moment.

\subsection{First moment}

For the SK model, the energy $E(\bss)$ is a Gaussian
random variable with density given by \eqref{g-SK}.
For $\alpha_n=O(n)$ and $\tilde\xi_n=o(1)$,
the first moment can therefore be approximated as
\begin{equation}
\label{SK.1stmoment-bd}
\begin{aligned}
\EX[Z_n(a_n^\ell,b_n^\ell)]
&=\frac {2^{n-1}}{\sqrt{2\pi n}}\int_{a_n^\ell}^{b_n^\ell}
e^{-\frac 1{2n}x^2}dx
\\
&=\frac {\gamma_\ell\tilde\xi_n2^{n-1}}{\sqrt{2\pi n}}
e^{-\frac 1{2n}\alpha_n^2}
\Bigl(1+O\Bigl(\frac{\alpha_n\tilde\xi_n}n\Bigr)
+O\Bigl(\frac{\tilde\xi_n^2}n\Bigr)
\Bigr)
\\
&=\gamma_\ell(1+o(1)).
\end{aligned}
\end{equation}
This proves the convergence of the first moment for
$\alpha_n\leq cn$ and $c<\sqrt {2\log 2}$.

\subsection{Families of linearly independent configurations}

When analyzing the factorization properties of the
joint distribution
of $E(\bss^{(1)})$, $\dots$, $E(\bss^{(k)})$, we will
want to use the representation
\eqref{SK-density-k}, which at a minimum, requires
that the covariance matrix $C$ defined in \eqref{C-SK}
is invertible.  The following Lemma~\ref{lem:SK-comb}
shows that a family of linearly independent configurations
leads to an invertible covariance matrix, and gives
a bound on the contribution of the families which are
not linearly independent.
It serves the same purpose as Lemma~\ref{lem:comb-bds} (2)
served for the {\npp}.  Note that statements  similar to those
in Lemma~\ref{lem:SK-comb} were proved in
\cite{bovier:kurkova:05} under the more restrictive condition
that  $\alpha_n=O(n^{-\eta})$ with $\eta<1$.

To state the lemma, we define
$\tilde R_{n,k}$ as
\begin{equation}
\label{SK-Error-def}
\tilde R_{n,k}=
\frac 1{2^k}
\sump_{\bss^{(1)},\cdots ,\bss^{(k)}}
 \PR\Bigl(E(\bss^{(j)})\in [a_n^{\ell(j)},b_n^{\ell(j)}]
  \text{ for all } j=1,\dots,k\Bigr),
\end{equation}
where the sum runs over pairwise distinct configurations
$\bss^{(1)},\dots, \bss^{(k)}$ which are linearly
dependent.

\begin{lemma}
\label{lem:SK-comb}
Let $k<\infty$.  Then there exists a constant $\epsilon_k>0$ such that
the following statements are true.

(1) If $\bss^{(1)}, \dots, \bss^{(k)}$ are linearly independent, the
matrix $C$  defined in \eqref{C-SK} is invertible, and
\begin{equation}
\label{SK-g-bd}
\tilde g^{(k)}(\bold x)\leq\Bigl(\frac{n}{2\pi}\Bigr)^{k/2}
e^{-\frac 12 (\bold x,C^{-1}\bold x)}.
\end{equation}

(2) If $\alpha_n\leq n\epsilon_k$, then
\begin{equation}
\tilde R_{n,k}\to 0\qquad\text{as}\qquad n\to\infty.
\end{equation}
\end{lemma}

\begin{proof}
(1) This statement is quite easy and must be known to most experts
in the field.  First,  we rewrite the matrix elements of $C$ as
$C_{ij}=nq(\boldsymbol\eta^{(i)},\boldsymbol\eta^{(j)})$, where
$\boldsymbol\eta^{(1)}$, $\dots$, $\boldsymbol\eta^{(k)}$ are
vectors in $\{-1,+1\}^{n^2}$.  Indeed, setting
$\eta_{r,s}^{(i)}=\sigma_s^{(i)}\sigma_r^{(i)}$, where $r,s=1,\dots
n$ and $i=1,\dots k$, we see that
$q(\boldsymbol\eta^{(i)},\boldsymbol\eta^{(j)})=
\bigl(q(\bss^{(i)},\bss^{(j)})\bigr)^2$, implying the above
representation for $C$.  Next we observe that that the linear
independence of $\bss^{(1)}, \dots, \bss^{(k)}$ implies linear
independence of $\boldsymbol\eta^{(1)}$, $\dots$,
$\boldsymbol\eta^{(k)}$, which in turn can easily be seen to imply
linear independence of the row vectors of the matrix with matrix
elements $q(\boldsymbol\eta^{(i)},\boldsymbol\eta^{(j)})$.  This
gives that $C$ is non-degenerate. But $C$ is a $k\times k$ matrix
with entries which are multiples of $n^{-1}$,  so if $\det C\neq 0$,
we must have $|\det C|\geq n^{-k}$.  This implies the bound
\eqref{SK-g-bd}.

(2) To prove (2), we decompose the sum in \eqref{SK-Error-def} according
to the rank of the matrix $M$ formed by the  vectors $\bss^{(1)}, \dots, \bss^{(k)}$.
Assume that the rank of $M$ is equal to
$u<k$. Reordering the vectors $\bss^{(1)}, \dots, \bss^{(k)}$, if
necessary, let us further assume that
$\bss^{(1)}, \dots, \bss^{(u)}$ are linearly independent.  With the help of
(1), we then
bound the probability on the right hand side of \eqref{SK-Error-def} by
\begin{equation}
\begin{aligned}
\PR\Bigl(&E(\bss^{(j)})\in [a_n^{\ell(j)},b_n^{\ell(j)}]
\text{ for all } j=1,\dots,k\Bigr)
\\
&\leq
   \PR\Bigl(E(\bss^{(j)})\in [a_n^{\ell(j)},b_n^{\ell(j)}]
  \text{ for all } j=1,\dots,u\Bigr)
\\
&\leq \Bigl(\frac{n}{2\pi}\Bigr)^{u/2}\prod_{j=1}^u(\tilde\xi_n\gamma_{\ell(j)})
=O((\sqrt n\tilde\xi_n)^u)
\end{aligned}
\end{equation}
To continue, we use the following two facts, proven, e.g.,
in \cite{borgs:chayes:pittel:01} (see also Lemma 3.9 in \cite{part1}:

\begin{enumerate}

\item If $\bss^{(1)}, \dots, \bss^{(k)}$
are pairwise distinct, the rank of $M$ is $u<k$ and
$\bss^{(1)}$, $\dots$, $\bss^{(u)}$ are linearly independent, then
$n_\bsd (\bss^{(1)}, \dots, \bss^{(u)})=0$
for at least one $\bsd\in\{-1,+1\}^u$.

\item Given $u<k$ linearly independent vectors $\bss^{(1)}, \dots, \bss^{(u)}$,
there are at most $2^{u(k-u)}$ ways to choose $\bss^{(1)}, \dots, \bss^{(k)}$
in such a way that the rank of $M$ is $u$.

\end{enumerate}
As a consequence, of (1), we have
\[
|\max_{\bsd}|n_\bsd (\bss^{(1)}, \dots, \bss^{(u)})-2^{-u}n|
\geq 2^{-u}n.
\]
Combined with Lemma~\ref{lem:comb-bds} (1) and the property (2) above,
we conclude that there are at most
$O(2^{nu}e^{-2^{-(2u+1)}n})=
O(2^{nu}e^{-2^{-(2k-1)}n})$ ways to choose $k$ pairwise distinct
configurations $\bss^{(1)}, \dots, \bss^{(k)}$ such that the rank of $M$
is $u$.  Using the fact that $(\sqrt n\tilde\xi_n)^u=
O(n^ue^{\frac u{2n}\alpha_n^2}2^{-nu})$, we immediately see that
for $\alpha_n\leq\epsilon_kn$ and $\epsilon_k$ sufficiently small,
the contribution of all configurations $\bss^{(1)}, \dots, \bss^{(k)}$
such that the rank of $M$ is smaller than $k$ decays exponentially in $n$,
so in particular it is $o(1)$.
\end{proof}

Consider now a family of linearly independent configurations
$\bss^{(1)},\dots,\bss^{(k)}$.  We claim that for such
a family and $\alpha_n\leq cn$ with $c<\sqrt {2\log 2}$,  we have
\begin{equation}
\label{SK-prob=dens}
\begin{aligned}
&\PR\Bigl(E(\bss^{(j)})\in [a_n^{\ell(j)},b_n^{\ell(j)}]
\text{ for all } j=1,\dots,k\Bigr)
\\
&\qquad=
\prod_{\ell=1}^m(\tilde\xi_n\gamma_\ell)^{k_\ell}
\frac 1{(2\pi)^{k/2}}\frac 1{(\det C)^{1/2}}
e^{-\frac 12(\boldsymbol\alpha, C^{-1}\boldsymbol\alpha)}
(1+o(1)),
\end{aligned}
\end{equation}
where
$\boldsymbol\alpha$ is the vector
$(\alpha_n,\dots,\alpha_n)\in\R^k$ and $C$ is the covariance
matrix defined in \eqref{C-SK}.  To prove this approximation,
we have to show that $(\bold x,C^{-1}\bold x)
=(\boldsymbol\alpha,C^{-1}\boldsymbol\alpha)=o(1)$
whenever $\bold x$ is a vector with $x_i=\alpha_n+O(\tilde\xi_n)$.
This in turn requires an upper bound on the inverse of $C$.  To
prove such a bound
we use that the matrix elements of $C$ are  bounded by
$n$, while $\det C$ is bounded from below by $n^{-k}$.
Using Cramer's rule, we conclude that the norm of $C^{-1}$ is
$O(n^{2k-1})$, which in turn implies that
$
(\bold x,C^{-1}\bold x)
=(\boldsymbol\alpha,C^{-1}\boldsymbol\alpha) +O(n^{2k-1}\alpha_n\tilde\xi_n)
+O(n^{2k-1}\tilde\xi_n^2)
$.
For $\alpha_n\leq cn$ with $c<\sqrt {2\log 2}$, the error term is $o(1)$,
as desired.

Assume that $\alpha_n\leq cn$ with $c<\min\{\sqrt {2\log 2},\epsilon_k\}$.
Using first the representation
\eqref{Zm-multi}, then
 Lemma~\ref{lem:SK-comb} and the bound \eqref{SK-prob=dens},
 and finally the explicit formula \eqref{tilde-xi-n}
for $\tilde\xi_n$, we
now approximate the $k^{\text{th}}$ factorial moment as
\begin{equation}
\begin{aligned}
\label{Zm-multi-SK}
\mathbb{E}
\Bigl[\prod_{\ell=1}^m
&(Z_n(a_n^{\ell},b_n^{\ell}))_{k_{\ell}}\Bigr]
=
\Bigl(\prod_{\ell=1}^m
\gamma_\ell^{k_\ell}\Bigr)
\\&
\times 2^{-nk}\sumpp_{\bss^{(1)} ,\dots, \bss^{(k)}}
\frac {n^{k/2}}{(\det C)^{1/2}}
 e^{\frac 1{2n}\|\boldsymbol\alpha\|_2^2}
 e^{-\frac 12(\boldsymbol\alpha, C^{-1}\boldsymbol\alpha)}
\Bigl(1+o(1)\Bigr)+o(1)
\end{aligned}
\end{equation}
where the sum goes over families of linearly independent configurations
$\bss^{(1)}$, $\dots$, $\bss^{(k)}$.

\subsection{Poisson convergence for
$\boldsymbol\alpha_{\bold n}\boldsymbol=
\bold o\boldsymbol(\bold n\boldsymbol)$}

In this section, we prove Theorem~\ref{thm:main1-SK}.
To this end, we again choose $\lambda_n$ in
such a way that $\alpha_n=o(\lambda_n\sqrt n)$,
$\lambda_n=o(\sqrt n)$ and $e^{-\lambda_n^2/2}$
decays faster than any power of $n$.
Recalling Lemmas~\ref{lem:comb-bds} (1) and (3), we may then
restrict the sum in \eqref{Zm-multi-SK} to a sum
over configurations with
$q_{\max}=O(\lambda_n/\sqrt n)$.
Expanding the inverse of $C$ as
$C^{-1}_{ij}=\frac 1n(\delta_{ij}+O(q_{\max}^2))$,
we then approximate $\det C$ as $\det C=n^k(1+o(1))$,
and $(\boldsymbol\alpha, C^{-1}\boldsymbol\alpha)$
as $(\boldsymbol\alpha, C^{-1}\boldsymbol\alpha)
=\|\boldsymbol\alpha\|_2^2+O(n^{-1}\alpha_n^2q_{\max}^2)$.
Using Lemma~\ref{lem:comb-bds} and Lemma~\ref{lem:SK-comb}
a second time to extend the sum over families
of configurations back
to a sum over all families of configurations
in $\{-1,+1\}^n$, the proof of Theorem~\ref{thm:main1-SK}
is therefore reduced to the proof of
the bound
\begin{equation}
\label{Zm-multi-SK-1}
2^{-nk}\sum_{\bss^{(1)} ,\dots, \bss^{(k)}}
e^{O(\frac 1n\alpha_n^2q_{\max}^2)}
=1+o(1).
\end{equation}
Since typical configurations lead to a maximal off-diagonal overlap
$q_{\max}$ of order $O(n^{-1/2})$, we expect that such a bound holds
as long as $\alpha_n=o(n)$. Lemma~\ref{lem:secterm} in
Section~\ref{sec:lemG1-proof} implies that this is indeed the case.

\subsection{Absence of Poisson convergence for
faster growing
$\boldsymbol\alpha_{\bold n}$}

In this section, we prove Theorem~\ref{thm:main2-SK}.
We start by proving the following lemma.

\begin{lemma}
\label{lem:SK-Zmbd}
Given a positive integer $k$,  there are constants
$\tilde\epsilon_k>0$ and $C_k<\infty$ such that
\begin{equation}
\label{Zm-multi-SK-bd}
\mathbb{E}
\Bigl[\prod_{\ell=1}^m
(Z_n(a_n^{\ell},b_n^{\ell}))_{k_{\ell}}\Bigr]
\leq
C_k\prod_{\ell=1}^m
\gamma_\ell^{k_\ell}+o(1)
\end{equation}
whenever
$\alpha_n\leq \tilde\epsilon_k n$,
$k=\sum_\ell k_\ell$ and $n$ is large enough.
\end{lemma}

\begin{proof}
Starting from the approximation \eqref{Zm-multi-SK},
we will again restrict ourselves to configurations
obeying the condition \eqref{const-lan}, but this
time we will choose $\lambda_n$ of the form
$\lambda_n=\sqrt {2kn}\tilde\epsilon_k$ where $\tilde\epsilon_k>0$
will be chosen in such a way that
$\tilde\epsilon_k<\sqrt{2\log 2}$ and
$\tilde\epsilon_k\leq \epsilon_k$.  Note that our choice of
$\lambda_n$  guarantees that
for
 $\alpha_n\leq n\tilde\epsilon_n$,
the sum over families of configurations violating the condition
\eqref{const-lan} decays exponentially with $n$. Note further that
the condition \eqref{const-lan} with $\lambda_n=\sqrt
{2kn}\tilde\epsilon_k$ guarantees that $q_{\max}^2\leq
2^{2k+1}{k}\tilde\epsilon_k^2$. Expanding $\det C$ as $\det
C=n^k(1+O(q_{\max}^2))$, we therefore have that $\det C\geq \frac 12
n^k$ provided $\tilde\epsilon_k$ is chosen small enough.

Let us finally write the matrix $C$ as $C=\frac 1n
(\mathbb{I}_k+A)$, where $\mathbb{I}_k$ is the identity matrix of
size $k$, and $A$ is the matrix with zero diagonal and off-diagonal
entries $q^2(\bss^{(i)},\bss^{(j)})$. Expanding $C^{-1}$ as
$C^{-1}=\frac 1n(\mathbb{I}_k-A+\frac{A^2}{1+A})$, we see that
\begin{equation}
(\boldsymbol\alpha,C^{-1}\boldsymbol\alpha)
\geq \frac 1n
\Bigl(\|\boldsymbol\alpha\|_2^2-(\boldsymbol\alpha,A\boldsymbol\alpha)
\Bigr)
\geq \frac 1n\Bigl(
\|\boldsymbol\alpha\|_2^2-
k^2q_{\max}^2\alpha_n^2\Bigr).
\end{equation}
Together with our lower bound on $\det C$, we see that the proof of
the lemma now reduces to the bound
\begin{equation}
\label{Zm-multi-SK-2}
2^{-nk}\sum_{\bss^{(1)} ,\dots, \bss^{(k)}}
\sqrt 2
 e^{\frac 1{2n}(k\alpha_nq_{\max})^2}
 \leq C_k.
\end{equation}
For $\tilde\epsilon_k$ small enough, this bound again follows from
Lemma~\ref{lem:secterm} in Section~\ref{sec:lemG1-proof}.
\end{proof}

Note that Lemma~\ref{lem:SK-Zmbd} implies statement (ii) of
Theorem~\ref{thm:main2-SK}, while the bound \eqref{SK.1stmoment-bd}
implies statement (i).  We are therefore left with the proof of
(iii), i.e., the statement that $\limsup_{n\to\infty}
\EX[(Z_n(a,b))_2]>\gamma^2$ when $\limsup\alpha_n/n>0$. Recalling
the representation, this in turn requires us to prove that
\begin{equation}
\label{SK.to-be-bounded0}
\limsup_{n\to\infty}
2^{-2n}\sumpp_{\bss^{(1)} , \bss^{(2)}}
\frac {n}{(\det C)^{1/2}}
 e^{\frac 1{2n}\|\boldsymbol\alpha\|_2^2}
 e^{-\frac 12(\boldsymbol\alpha, C^{-1}\boldsymbol\alpha)}
 >1.
\end{equation}
Let $q=q(\bss^{(1)} ,\bss^{(2)})$
be the off-diagonal overlap of the two configurations
$\bss^{(1)} $ and $\bss^{(2)}$.  Since $C$ is now just a $2\times 2$ matrix,
both its inverse and its determinant can be easily calculated,
giving $\det C=n^2(1-q^4)$, and, using that
$\|\boldsymbol\alpha\|_2^2 = k \alpha_n^2 = 2 \alpha_n^2$,
\begin{equation}
(\boldsymbol\alpha, C^{-1}\boldsymbol\alpha)=
\frac 1{n(1-q^4)}(2\alpha_n^2-2q^2\alpha_n^2)
=\frac{2\alpha_n^2}{n(1+q^2)}
=\frac{2\alpha_n^2}n(1-\frac{q^2}{1+q^2}).
\end{equation}
As a consequence, we will have to prove a lower bound on the
expression
\begin{equation}
\label{SK.to-be-bounded}
2^{-2n}\sumpp_{\bss^{(1)} , \bss^{(2)}}
\frac {1}{\sqrt {1-q^4}}
 e^{\frac{\alpha_n^2}n\frac{q^2}{1+q^2}}.
\end{equation}
Taking subsequence, if necessary, let us assume that $\alpha_n/n$
converges to some $\kappa$, with
$0<\kappa<\min\{\epsilon_2,\sqrt {2\log 2}\}$. We then bound the
sum $\sumpp$ from below by restricting it to all configurations for
which $|q|\leq 2\kappa$. Under this restriction,  the summand is
bounded from below by $\frac 1{\sqrt{1-16\kappa^4}}e^{\beta_n
q^2}$, with $\beta_n=\alpha_n^2/(n(1+4\kappa^2))$. Observing that
$q^2\leq 1$ we may then use Lemma~\ref{lem:comb-bds} (1) and (3) to
extend the sum back to a sum which runs over all configurations
$\bss^{(1)}$ and $\bss^{(2)}$, leading to the lower bound
\begin{equation}
\label{SK.to-be-bounded1}
2^{-2n}\sumpp_{\bss^{(1)} , \bss^{(2)}}
\frac {1}{\sqrt {1-q^4}}
 e^{\frac{\alpha_n^2}n\frac{q^2}{1+q^2}}
 \geq
\frac {2^{-2n}}{\sqrt {1-16\kappa^4}}
\sum_{\bss^{(1)} , \bss^{(2)}}
 e^{\beta_n q^2}+o(1).
\end{equation}
Let $\EX_2$ denote expectations with respect to the uniform
measure over families
of configurations $\bss^{(1)} , \bss^{(2)}\in\{-1,+1\}^n$.
Using H\"older and the fact that $\EX_2[q^2]=\frac 1n$, we then
lower bound the right hand side by
\begin{equation}
\label{SK.to-be-bounded2}
\begin{aligned}
\frac {2^{-2n}}{\sqrt {1-16\kappa^4}}
&\sum_{\bss^{(1)} , \bss^{(2)}}
 e^{\beta_n q^2}+o(1)
\\
 &\geq\frac {e^{\beta_n/n}}{\sqrt {1-16\kappa^4}}+o(1)
=\frac 1{\sqrt {1-16\kappa^4}}
 \exp\Bigl(\frac{\kappa^2}{1+4\kappa^2}\Bigr)+o(1).
 \end{aligned}
\end{equation}
For $\kappa$ sufficiently small, the right hand side is
asymptotically larger than $1$, proving the desired lower
bound \eqref{SK.to-be-bounded0}, which completes the proof
of Theorem~\ref{thm:main2-SK}.

\section{Auxiliary results}
\label{sec:Aux}

\subsection{Proof of Lemma~\ref{lem:sum-factor}}

 \begin{proof}[Proof of Lemma~\ref{lem:sum-factor}]
We will have to prove that
\begin{equation}
\label{sum-factor0}
 2^{-nk}
\sum_{\bss^{(1)},\dots,\bss^{(k)}}
e^{f(\bss^{(1)},\dots,\bss^{(k)})}
=1+o(1).
\end{equation}
Recalling the assumption $|f(\bss^{(1)},\dots,\bss^{(k)})| \leq
{c}\alpha^2_nq_{\max}$, let $\theta_n =o(n)$ be  a sequence of
positive integers such that $\theta_n \to \infty $ as $n \to \infty$
and $\alpha_n^2 \theta_n = o(\sqrt n)$.

We now split the sum  on the left hand side of \eqref{sum-factor0}
into two parts:
the sum over all families of configurations
$\bss^{(1)},\dots,\bss^{(k)}$ that satisfy \eqref{const-lan} with
$\theta_n$ taking the place of $\lambda_n$ and the sum over the
configurations that violate the bound \eqref{const-lan}, again with
$\theta_n$ replacing $\lambda_n$.

Consider the first sum. By Lemma~\ref{lem:comb-bds}, the number of
terms in this sum is bounded between $2^{nk}(
1-2^{k+1}e^{-\frac{\theta_n^2}{2}})$ and $2^{nk}$. For all these
configurations we know from Lemma \ref{lem:comb-bds} that $q_{max}
\leq 2^k \frac{\theta_n}{\sqrt{n}}$  and hence
$|f(\bss^{(1)},\dots,\bss^{(k)})| \leq c 2^k \alpha_n^2
\frac{\theta_n}{\sqrt{n}} = o(1)$ as $n \to \infty$. Using the fact
that $\theta_n \to \infty$ as $n \to \infty$ we obtain that the
contribution from the first summation is $1 + o(1)$ as $n \to
\infty$.

Thus, to establish Lemma \ref{lem:sum-factor} it suffices to show
that the contribution from the second sum is $o(1)$ as $n \to
\infty$. To this  end we partition the interval $[ \theta_n, n]$
into  subintervals $[\kappa_i,\kappa_{i+1}]$ of length one. Now
consider the families of configurations
$\bss^{(1)},\dots,\bss^{(k)}$ that satisfy \eqref{const-lan} when
$\lambda_n$ is replaced by $\kappa_{i+1}$ but violate  it if
$\lambda_n$ is replaced by $\kappa_i$. It is easy to see that  for
$n$ large enough, the contribution from these terms is bounded by
\[
2^{k+1} e^{-\kappa_i^2/2}e^{c\alpha_n^22^k\frac{\kappa_{i+1}}{\sqrt n}}
=
2^{k+1} e^{-\kappa_i^2/2}e^{o(\kappa_i)}
\leq e^{-\kappa_i^2/4}.
\]
Adding up these error terms, we obtain that the second sum is of
order $O(e^{-\theta_n^2/4})=o(1)$.  This completes the proof of
Lemma \ref{lem:sum-factor}.
\end{proof}

\subsection{Proof of Lemma~\ref{lem:G1}}
\label{sec:lemG1-proof}

The proof of Lemma~\ref{lem:G1} will be based on the following
lemma.

\begin{lemma}
\label{lem:secterm} Let $k$ be a positive integer, and let
$\beta<\frac n{k(k-1)}$ be a non-negative real. If $|g(\bss^{(1)}
,\dots, \bss^{(k)})| \leq \beta q_{\max}^2$ for all families of
configurations $\bss^{(1)},\dots,\bss^{(k)}$, then
\begin{equation}
\Bigl(1-\frac {k(k-1)}2\frac \beta n\Bigr)
\leq 2^{-nk}
\sum_{\bss^{(1)},\dots, \bss^{(k)}}
 e^{g(\bss^{(1)} ,\dots, \bss^{(k)})}
 \leq \Bigl(1-k(k-1)\frac\beta n\Bigr)^{-1/2}
\end{equation}
\end{lemma}

\begin{proof}
Let $\EX_k[\cdot]$ denote expectations with respect to the uniform
measure over all families of configurations
$\bss^{(1)},\dots,\bss^{(k)}\in\{-1,+1\}^n$.  With the help of
Jensen's inequality, we then immediately obtain the lower bound
\begin{equation}
\label{aux.low}
\begin{aligned}
\EX_k[e^g]&\geq\exp(-\beta\EX_k[q_{\max}^2])
\geq \exp\Bigl(-\beta\sum_{i<j}\EX_k[q_{ij}^2]\Bigr)
\\
&=\exp\Bigl(-\beta \frac{k(k-1)}2\frac 1n\Bigr) \geq 1-\beta
\frac{k(k-1)}2\frac 1n .
\end{aligned}
\end{equation}
To obtain an upper bound, we first use
H\"older's inequality to obtain the estimate
\begin{equation}
\label{aux.Hoelder} \EX_k[e^g]\leq
\EX_k[e^{\sum_{i<j}q_{ij}^2}] \leq
\prod_{i<j}\EX_k[e^{K\beta
q_{ij}^2}]^{1/K} \leq \max_{i<j}\EX_k[e^{K\beta q_{ij}^2}]
\end{equation}
with $K=k(k-1)/2$.  Next we rewrite the expectation on
the right hand side
as
\begin{equation}
\label{aux.int}
\EX_k[e^{K\beta q_{ij}^2}]=
\frac 1{\sqrt {2\pi}}\int_{-\infty}^\infty e^{-\frac{x^2}2}
\EX_k[e^{\sqrt {2K\beta} q_{ij}x}] dx.
\end{equation}
But now we can calculate the expectation exactly.
Together with the inequality $\cosh (y)\leq \exp(y^2/2)$,
this leads
to the estimate
\[
\EX_k[e^{\sqrt {2K\beta} q_{ij}x}] = \Bigl(\cosh(\sqrt
{2K\beta}xn^{-1})\Bigr)^n \leq\exp\Bigl(K\beta x^2n^{-1}\Bigr).
\]
Inserting this into \eqref{aux.int} and \eqref{aux.Hoelder}, we have
\[
\EX_k[e^g]
\leq\frac 1{\sqrt{2\pi}}\int_{-\infty}^\infty
e^{-(1-2K\beta n^{-1})\frac{x^2}2} dx
=\frac 1{\sqrt{1-2K\beta n^{-1}}}
=\frac 1{\sqrt{1-k(k-1)\beta n^{-1}}},
\]
the desired upper bound.
\end{proof}

Having established the above lemma, we are now ready
to prove Lemma~\ref{lem:G1}.

\begin{proof}[Proof of Lemma~\ref{lem:G1}]
Let $f_1(\bss^{(1)} ,\dots, \bss^{(k)}) =\sum_{i\neq j} \tilde
q_{ij}$ and $f=f_1+R$, and  define
\[
F(\beta) = \EX_k[e^{\beta f}],
~~ F_1(\beta) = \EX_k[e^{\beta f_{ 1}}],
~\mbox{and}
~~ F_2(\beta) = \EX_k[e^{{\beta R}}]
 .
\]
Since (by \eqref{lemG.2})
\begin{equation}
\label{F.1-bd} F_1(\beta_n)= \exp\Bigl(\frac{k(k-1)}n\beta_n^2+
O(\beta_n^3n^{-2}) \Bigr),
\end{equation}
we only have to prove that $F(\beta_n) = F_1(\beta_n) e^{
O(\beta_n^3 n^{-2})+o(1)}$ whenever $\beta_n = o(n)$.

Let $\epsilon_n = 2k(k-1)C\frac{\beta_n}{n}$  where $C$ is the
constant from Lemma~\ref{lem:G1}.  Applying H\"{o}lder's inequality
to  $\EX_k[e^f]$ we obtain
\[ F(\beta_n) \leq F_1(\beta_n(1+\epsilon_n))^{\frac 1{1+\epsilon_n}}
F_2(\beta_n\frac{1+\epsilon_n}{\epsilon_n})^{\frac{\epsilon_n}{1+\epsilon_n}}
\]
and hence
\[
  \frac{F(\beta_n)}{F_1(\beta_n)}
  \leq
  \frac{ F_1(\beta_n(1+\epsilon_n))^{\frac 1{1+\epsilon_n}}}{F_1(\beta_n)}
  F_2(\beta_n\frac{1+\epsilon}{\epsilon})^{\frac{\epsilon_n}{1+\epsilon_n}}.
\]
Using equation \eqref{F.1-bd} one obtains
\[
  \frac{ F_1(\beta_n(1+\epsilon_n))^{\frac 1{1+\epsilon_n}}}{F_1(\beta_n)} =
  \exp( \beta_n^2\epsilon_nn^{-1} + O(\beta_n^3/n^2)) = \exp(O(\beta_n^3/n^2)).
\]
Recalling that $|R| \leq Cq_{\max}^2$,
we would like to
use
Lemma \ref{lem:secterm} to bound the second factor.
To this end, we need to guarantee that
$k(k-1)\frac{1+\epsilon_n}{\epsilon_n}C\frac{\beta_n}n<1$.
By our choice of $\epsilon_n$
this will be the case as soon as $n$ is large enough to
ensure that $\epsilon_n<1$.  For $n$ large enough,
we therefore have
\[
  F_2(\beta_n\frac{1+\epsilon}{\epsilon})^{\frac{\epsilon}{1+\epsilon}}
\leq \Bigl(
1-k(k-1)\frac{1+\epsilon_n}{\epsilon_n}C\frac{\beta_n}n
\Bigr)^{-\frac{\epsilon_n}{2(1+\epsilon_n)}}
=\exp(O(\beta_n/n)).
\]
Putting everything together, we get
\[
  \frac{F(\beta_n)}{F_1(\beta_n)}
  \leq
  \exp\Bigl(O(\beta_n^3/n^2)+O(\beta_n n^{-1})\Bigr)
  =\exp\Bigl(O(\beta_n^3/n^2)+o(1)\Bigr).
\]

Applying H\"{o}lder's inequality to
$\EX_k[e^{(1+\epsilon_n)^{-1}\beta_n (f- R)}]$ we obtain
that
\[
F_1\Bigl( \frac{\beta_n}{1+\epsilon_n}\Bigr)^{1+\epsilon_n} \leq
F(\beta_n) F_2\Bigl(-\frac{\beta_n}{\epsilon_n}\Bigr)^{\epsilon_n}.
\]
This implies that
\[
\frac{F(\beta_n)}{F_1(\beta_n)} \geq
\frac{F_1((1+\epsilon_n)^{-1}\beta_n)^{1+\epsilon_n}}{F_1(\beta_n)}
\frac{1}{F_2(-\beta_n\frac{1}{\epsilon_n})^{\epsilon_n}}.
\]
Proceeding along similar lines  as for the upper bound, we obtain
that
\[
  \frac{F(\beta_n)}{F_1(\beta_n)}
  \geq
  \exp\Bigl(O(\beta_n^3/n^2)+o(1)\Bigr).
\]

Combining the two applications of H\"{o}lder's inequality we finally obtain that
\[
  \frac{F(\beta_n)}{F_1(\beta_n)}
  =
  \exp\Bigl(O(\beta_n^3/n^2)+o(1)\Bigr).
\]
This completes the proof of the lemma.
\end{proof}

\section{Summary and Outlook}
\label{sec:Discuss}

\subsection{Summary of Results}

In this paper, we considered the local  REM conjecture of Bauke,
Franz and Mertens,
for both the \npp~and the SK spin glass model.
For the \npp~we showed
that the local REM conjecture
holds for energy scales $\alpha_n$ of order $o(n^{1/4})$,
and fails if
$\alpha_n$ grows like $\kappa n^{1/4}$ with $\kappa>0$.
For the SK model we established a similar threshold,
showing that the local REM conjecture
holds for energies of order
$o(n)$, and fails if the energies grow like
$\kappa n$ with $\kappa >0$ sufficiently small.

Although we believe that the local REM conjecture also fails for
still faster growing energy scales, our analysis did not allow
us to make this rigorous since we could not exclude that
the moments of the energy spectrum diverge, while the
spectrum itself undergoes a re-entrance transition and
converges again to Poisson for faster growing energy scales.

\subsection{Proof Strategy}
Before discussing possible extensions, let us recall our
proof strategy.  Both the proof of Poisson convergence, and the
proof that Poisson convergence fails after a certain point,
relied on a precise asymptotic analysis of the factorial
moments.

For the purposes of this discussion, let us restrict
to the one-dimensional factorial moments $\EX[(Z_n(a_n,b_n))_k]$
of the number of points in the energy spectrum between
some
$a_n$ and $b_n$ in the vicinity of $\alpha_n$.
We expressed these factorial moments in terms
of the probability that the energies
$E(\bss^{(1)}),\dots,E(\bss^{(k)})$
of $k$ pairwise distinct
configurations $\bss^{(1)},\dots,\bss^{(k)}$ all lie
in the interval $[a_n,b_n]$, see \eqref{Zm-multi}.

The  technical meat of our
proof then consisted of two steps: a bound on the sum over
``atypical configurations'' (see Lemmas~\ref{lem:comb-bds} and
\ref{lem:SK-comb}), and a proof that for typical
configurations, the probability that the energies of
$\bss^{(1)},\dots,\bss^{(k)}$ all lie
in the interval $[a,b]$ is asymptotically equal to the
product of the probabilities $\PR(E(\bss^{(i)}\in [a_n,b_n]))$.

For both the {\npp} and the SK model, the value of the threshold
can already be understood by considering the factorization
properties of typical configurations, see \eqref{Zm-multi-2}.
Taking, e.g., the case
$k=2$, our results say that for typical configurations in the {\npp}
with Gaussian noise, we have factorization up to a factor
\begin{equation}
e^{\alpha_n^2(q+O(q^2))}(1+o(1)),
\end{equation}
where $q$ is the overlap between
the two configurations.  Since the overlap between typical configurations
is of order $n^{-1/2}$, we obtain factorization if and only if
$\alpha_n=o(n^{1/4})$.  To obtain the same result for the {\npp}
with general distribution was much more work, since it required
establishing a large deviation density estimate for $k$ {\it a priori}
highly dependent variables, but the heuristic for the threshold
of $n^{1/4}$ is still the same.

By contrast, the threshold for the SK model is easier to establish than
that for the {\npp}.  The main reason is the restriction to Gaussian noise,
which made the proof of a large deviation density estimate unnecessary,
in addition to simplifying the proof of the bound on atypical
configurations.  But on a heuristic level, there is not much of a difference:
now we obtain factorization up to multiplicative factor of
\begin{equation}
e^{\frac{\alpha_n^2}n(q^2+O(q^4))}(1+o(1)),
\end{equation}
giving the threshold of $\alpha_n=o(n)$ for Poisson convergence.

\subsection{$\bold p$-Spin models}

In the physics literature, one often considers a generalization of the
SK model which involves interactions between $p$ different spins, instead
of the two-body interaction of the standard SK model.  For our purpose,
these $p$-spin SK models are best defined as Gaussian fields indexed
by the spin configurations $\bss\in\{-1,+1\}^n$.  Recalling that
a Gaussian field is uniquely defined by its mean and covariance matrix,
we then define the  $p$-spin SK-Hamiltonian $H^{(p)}(\bss)$ as the Gaussian
field with mean $0$ and covariance matrix
$\EX[H^{(p)}(\bss) H^{(p)}(\tilde\bss)]=n q(\bss,\tilde\bss)^p$
where $p=1,2,\dots$,
and $q(\cdot,\cdot)$ is the overlap defined in \eqref{overlap}.
Note that the energy of the {\npp} with Gaussian noise is nothing
but the absolute value of the $p=1$ SK Hamiltonian divided by $\sqrt n$.
Up to a rescaling by $\sqrt n$, the energy spectrum of the {\npp}
with Gaussian noise is
therefore identical to the positive energy spectrum of the $p=1$ SK model.

It was shown in \cite{bovier:kurkova:05} that the local REM conjecture
holds for $p=1$ if $\alpha_n=O(n^{3/4-\epsilon})$ for some
$\epsilon>0$, and for $p\geq 2$ if $\alpha_n=O(n^{1-\epsilon})$.
For $p=1,2$ our results establish a little bit more, namely the validity
of the REM conjecture for $\alpha_n=o(n^{3/4})$ if $p=1$, and for
$\alpha_n=o(n)$ if $p=2$.  More importantly, our results prove that
these are actually the thresholds for the validity of the REM conjecture.

This raises the question about the true threshold for $p\geq 3$.
Starting  with the factorization properties for typical
configurations, one easily sees that the joint density
of $H(\bss^{(1)}),\dots,H(\bss^{(k)})$ is again given by a formula
of the form
\eqref{SK-density-k}, where $C$ is now the matrix with matrix element
$C_{ij}=n(q(\bss^{(i)},\bss^{(j)}))^p$.  This then leads to factorization
up to a multiplicative error term
\begin{equation}
e^{\frac{\alpha_n^2}nO(q_{\max}^p)}(1+o(1)),
\end{equation}
where $q_{\max}$ is the maximal off-diagonal overlap of
$\bss^{(1)},\dots,\bss^{(k)}$.  If $\alpha_n=O(n)$,
this gives factorization for typical configurations
as long as $p>2$.  But unfortunately, our control
over atypical configurations is not good enough
to allow a proof of the REM conjecture for energies
that grow that fast.
Indeed, it is easy to see
that Lemma~\ref{lem:SK-comb} can be generalized to
$p>2$; but an application of the lemma to control
the error for the $k^\text{th}$ factorial moment
requires the
condition $\alpha_n\leq n\epsilon_k$ where
$\epsilon_k$ is a positive constant that goes to zero as $k\to\infty$.

We therefore see that our methods can easily be used to
prove the local REM conjecture for $\alpha_n=o(n^{3/4})$ and
$p=1$ as well as $\alpha_n=o(n)$ and all $p\geq 2$,
but they are not strong enough to answer the
question whether this is the actual threshold for $p\geq 3$, or
whether the REM conjecture remains
valid if $p\geq 3$ and $\alpha_n$ grows like
$n\kappa$ for some small $\kappa>0$.

An even more challenging question is the question of what happens
after the local REM conjecture fails.  This question seems quite
hard, and so far it has only been answered for the hierarchical
GREM-model.  For this model it
has been shown \cite{bovier:kurkova:05b,bovier:kurkova:05c}
that the
suitably rescaled
energy spectrum converges to a mixed Poisson process with density
given in terms of  a  Poisson cascades on $\R^\ell$, where the
dimension $\ell$ becomes larger and larger as
$\kappa$ passes through an infinite series of thresholds,
the first threshold being the value where the local REM conjecture
fails for the GREM.

\subsection{A Simple Heuristic}

Returning now to the {\npp}, we note that the rigorous
moment analysis of the threshold is somewhat unintuitive
and rather involved.  Alternatively, let us present a simple heuristic
to explain the threshold scale
$n^{1/4}$.
To this end it is useful to introduce the gauge invariant magnetization
$$
M(\bss)
= \frac{1}{n} \sum \sigma_i  \text{sgn} X_i,
$$
and consider the joint distribution of $\widetilde M(\bss)
= \sqrt{n} M(\bss)$ and the ``signed''
energy $H(\bss)$ introduced in \eqref{H-def} with, as usual,
$X_1, \dots, X_n$ chosen i.i.d.~with density $\rho$, and
$\bss$ chosen uniformly at random.

The covariance matrix
$C$ of $\widetilde M(\bss)$ and $H(\bss)$ is easily calculated
to be
$$
C=\left[
\begin{matrix}
1&\mu\\
\mu&1
\end{matrix}
\right]
$$
where $\mu = \EX[|X_1|] < 1$, and its inverse $C^{-1}$ is given by
$$
C^{-1}=\frac 1{1-\mu^2}\left[
\begin{matrix}
1&-\mu\\
-\mu&1
\end{matrix}
\right].
$$
Assuming, for the moment, that the joint density of $\widetilde M(\bss)$
and  $H(\bss)$ obeys a local limit theorem, we now approximate this
density by the Gaussian density
\begin{equation}
\label{locallim}
\begin{aligned}
g(H,\widetilde M)
&= \frac {1}{2 \pi \sqrt{\det C}}
e^{-\frac 12(H^2C_{11}^{-1}+\widetilde M^2 C_{22}^{-1} +2H\widetilde M C_{12})}
\\
&= \frac {1}{2 \pi \sqrt{\det C}}
e^{-\frac 1{2(1-\mu^2)}(H^2+\widetilde M^2 -\mu H\widetilde M )}
\\
&= \frac {1}{2 \pi \sqrt{\det C}}
e^{-\frac 12 H^2}
e^{-\frac 1{2(1-\mu^2)}(\widetilde M -\mu H )^2}.
\end{aligned}
\end{equation}
In this approximation, the distribution of $\widetilde M$ conditioned
on $H = \alpha$ is therefore Gaussian with mean $\alpha\mu$ and
covariance $1-\mu^2$, implying in particular that the expectation
of $M$ is equal to $\alpha\mu/\sqrt{n}$.

Consider now two configurations $\bss, \bss'$, both chosen
uniformly at random among all configurations with magnetization
$M$.  Then the expected overlap between $\bss$ and $\bss'$
is $M^2(1+ o(1))$.

Finally, consider two configurations $\bss, \bss'$ both chosen
uniformly at random among all configurations with signed
energies in the range $H \in [\alpha, \alpha + d\alpha]$, for
some small $d\alpha$.
Since the energies of $\bss$ and $\bss'$ are correlated, it
follows that their magnetizations are also correlated.
Under the assumptions that (1) the joint distribution
$g(H,\widetilde M)$ obeys
a local limit theorem, as in \eqref{locallim}, and (2) the
{\it only} correlation between these configurations is due
to the correlation in their magnetizations, it would follow
that the expected overlap $\EX[q(\bss, \bss')]$ is given by
$M^2(1+ o(1)) = \alpha^2\mu^2/n (1+ o(1))$.

Note that the above heuristic focuses on the ``signed
energy'' $H(\bss)$ rather than the true energy $E(\bss)
= |H(\bss)|$.  Conditioning instead on
$E(\bss), E(\bss') \in [\alpha, \alpha + d\alpha]$,
the above heuristic therefore suggests a bimodal
distribution of $q(\bss, \bss')$ with peaks at
$\pm \alpha^2\mu^2/n (1+ o(1))$.

Now recall that the local REM conjecture says that the
overlap, rescaled by $\sqrt{n}$, is asymptotically normal.
However, our
above heuristic says that $\sqrt{n}q(\bss, \bss')$ cannot be
asymptotically normal for $\alpha$ growing like $n^{1/4}$
or faster, in
agreement with our rigorous results.  Thus this heuristic
correctly predicts the scale at which the REM conjecture
breaks down, and suggests that the breakdown could be due to
correlations in the magnetization of configurations with
similar energies of scale $n^{1/4}$ or greater.

A detailed calculation, however, suggests
that we exercise some caution in the application of this
heuristic.  Although the heuristic correctly predicts the
scale of the threshold, and the double peak structure of
the overlap at the threshold, it does not predict the
correct position of the peaks when $\alpha = \kappa n^{1/4}$.
Indeed, for this scaling, we rigorously showed that the rescaled
overlap distribution converges to a convex combination of
two Gaussians centered at $\pm \kappa^2$, in contrast to the
heuristic prediction of $\pm \kappa^2 \mu^2$.  This, in turn,
suggests that there are additional correlations besides those
induced by the magnetization.

\subsection{Algorithmic Consequences}

Over twenty years ago, Karmarkar and Karp \cite{KK82}
gave a linear time algorithm for a
suboptimal solution of the number partitioning problem.  For
i.i.d.~weights with densities of the form studied in the current paper,
the typical energy $E_{KK}$ of the KK  solution is of order
$n^{-\theta(\log n)}$, \cite{KK82,Yak96},
while the minimal energy is known to be much smaller
\cite{KKLO86,Lue98,borgs:chayes:pittel:01}, namely of
order $2^{-\theta(n)}$.

This raises the question of whether one can do better than the
KK solution -- a question which has received much study due to the
numerous applications of the number partitioning problem.
This work has led to many different heuristics,
but to our knowledge no algorithm with {\it guaranteed} performance
significantly better than KK has emerged.
In the absence of a good global alternative to KK, one might try to base an
improved solution of the problem on a local search algorithm starting from
the KK solution.  But such an approach is unlikely to produce better results,
as the following argument shows.

Consider the random {\npp} as defined in this
paper, and let $\bss$ be a partition with energy $E(\bss)$ of the order
of $E_{KK}$, i.e., $E(\bss)=n^{-\theta(\log n)}$.  Let $\tilde\bss$
be a local perturbation of $\bss$, i.e., let $\tilde\bss$ be a
configuration such that $\bss$ and $\tilde\bss$ differ on a small subset
$K\subset\{1,\dots,n\}$, with
$k=|K|$ bounded uniformly in $n$.
The signed energies $H(\bss)$ and $H(\tilde\bss)$
then
differ by $n^{-1/2}\Delta_K{(\bss_K)}$, where
$\Delta_K{(\bss_K)}$ is the random variable
\[
\Delta_K{(\bss_K)}=2\sum_{i\in K}\sigma_iX_i
\]
{and $\bss_K$ is the restriction of $\bss$ to $K$.}
Under mild assumptions on the
probability density $\rho$ of the weights $X_1,\dots,X_n$,
the density of $\Delta_K{(\bss_K)}$ is a
continuous function
near zero, and the
probability that $|\Delta_K|\leq \epsilon$ for some small $\epsilon$
is of order $\theta(\epsilon)$ {with the constants implicit in
the $\theta$
symbol depending only on $k$.

Obviously,  any local improvement algorithm that changes exactly $k$
bits will lead to a change in the unsigned energies that is bounded
from below by
$$
\delta_k^{(-)}=n^{-1/2}\min_{\bes K: |K|=k\es}\min_{\bss_K}|\Delta_K(\bss_K)|
.
$$
Taking into account that} there
are only $\binom{n}{k}\leq n^k$ possible choices for $K$,
{we see that the probability that
$\delta_k^{(-)}\leq \epsilon$ is bounded by
$O(\epsilon n^{k+1/2})$.
We conclude that with high probability,
$\delta_k^{(-)}$ is at least $\theta(n^{-1/2-k})$, }
much larger than the energy
$n^{-\theta(\log n)}$
of our starting configuration $\bss$.
Thus any  local improvement algorithm that changes
$k$ bits moves us with high probability
far away from the starting configuration
with energy of order $n^{-\theta(\log n)}$.

Note that this simple argument does not use very much; in particular, it is
not related to REM conjecture, which suggests a much deeper reason for the
apparent difficulty of the {\npp}.  Indeed, applying our local REM Theorem
to $\alpha=E_{KK}$, it says that in the vicinity of $E_{KK}$, the energy behaves
like a random cost function of $2^{n-1}$ independent random variables.  If the
energy of the {\npp} were truly a random cost function of  $2^{n-1}$ independent random variables,
this would imply that there is no algorithm faster than exhaustive search, implying
a running time exponential in $n$.
But of course, we have not come near to proving anything
as strong as this.

In fact, even on a non-rigorous level some caution is required when applying the
above heuristic.  Indeed, only $n$ linear independent configurations $\bss^{(1)}$,
$\dots$, $\bss^{(n)}$ are needed to completely determine the random variables
$X_1,\dots, X_n$ from the energies $E(\bss^{(1)}),\dots,E(\bss^{(n)})$, implying that
the energy spectrum lies in a subspace of dimension $n$, not $2^n$.
Still, our REM Theorem proves that the energy spectrum behaves locally
like that of a random cost function, suggesting several possible conjectures.

The first conjecture is best described in an oracle setting, where the oracle, $O$,
keeps the $n$ weights $X_1,\dots, X_n$ secret from the algorithm $A$.  The algorithm
is given the KK-solution $\bss^{(KK)}$ and its energy $E_{KK}$, and successively
asks the oracle for the energy of some configurations
$\bss^{(1)},\dots,\bss^{(m)}$, where $m$ is bounded.
Given this information, the
algorithm then calculates a new approximation $\tilde\bss\neq \pm\bss^{(KK)}$.
Given our REM Theorem,
we conjecture that with high probability (tending to one as $n\to\infty$),
the energy of the new configuration $\tilde\bss$ is much larger than
$E_{KK}$; in fact, we conjecture that with high probability
$E(\tilde\bss)/E_{KK}\to\infty$ as $n\to\infty$.

The second conjecture gives the algorithm much more input, namely the
$\ell$ first energies above a threshold $\alpha$
and the configurations corresponding to these energies.  The task of the algorithm
is now to find a new partition  $\tilde\bss$ whose energy is as near possible
to $\alpha$, which we assume to grow only slowly with $n$ (say like
$o(n^{-1/4})$, to stay in the realm of our Theorem \ref{thm:growing}).
 Again, the algorithm has no access to the original weights,
but may ask the oracle $O$ for the energies of $m$ additional configurations
$\bss^{(1)},\dots,\bss^{(m)}$, adapting the next question to the
answer of the preceding ones.  For $m$ and $\ell$  bounded uniformly in $n$,
we then conjecture that with high probability
the algorithm produces a configuration $\tilde\bss$ with
$(E(\tilde\bss)-\alpha)\xi_n^{-1}\to\infty$, while the actual value
of the next configuration above $\alpha$ is with high probability
equal to $\alpha+O(\xi_n)$ by our Theorem~\ref{thm:growing}.

One finally might want to consider the above conjectures in
a setting where the oracle
only reveals the relative order of the energies $E(\bss^{(1)}),\dots,E(\bss^{(m)})$,
but keeps the numerical values of $E(\bss^{(1)}),\dots,E(\bss^{(m)})$
secret.  In such a setting, there is no {\it a priori} reason to assume
that $m<n$.  Instead, it seems reasonable to conjecture inapproximability
results for values of $m$ and $\ell$ that are polynomial in $n$.

\bigskip
{\em Acknowledgement:}
We thank Keith Ball, Amir Dembo, Laci Lovasz, Yuval Perez, David Wilson and Avi
Widgerson for useful discussions. S.M.~was supported in part by the German
Science Council (grant
\mbox{ME2044/1-1}).


\end{document}